\documentclass[manuscript]{acmart}
\usepackage{csquotes}
\usepackage{subcaption}

\AtBeginDocument{%
  \providecommand\BibTeX{{%
    \normalfont B\kern-0.5em{\scshape i\kern-0.25em b}\kern-0.8em\TeX}}}

\setcopyright{acmcopyright}
\copyrightyear{2023}
\acmYear{2023}
\acmDOI{XXXXXXX.XXXXXXX}

\acmJournal{CSUR}
\acmVolume{37}
\acmNumber{4}
\acmArticle{111}
\acmMonth{8}

\begin{document}

\title{Mouse Dynamics Behavioral Biometrics: A Survey}

\author{Simon Khan}

\email{shkhan@clarkson.edu}
\affiliation{%
  \institution{Clarkson University}
  \streetaddress{8 Clarkson Avenue}
  \city{Potsdam}
  \state{NY}
  \country{USA}
  \postcode{13699}
}
\author{Charles Devlen}
\email{devlencm@clarkson.edu}
\affiliation{%
  \institution{Clarkson University}
  \streetaddress{8 Clarkson Avenue}
  \city{Potsdam}
  \state{NY}
  \country{USA}
  \postcode{13699}
}
\author{Michael Manno}
\email{mannom@clarkson.edu}
\affiliation{%
  \institution{Clarkson University}
  \streetaddress{8 Clarkson Avenue}
  \city{Potsdam}
  \state{NY}
  \country{USA}
  \postcode{13699}
}
\author{Daqing Hou}
\orcid{0000-0001-8401-7157}
\email{dhou@clarkson.edu}
\affiliation{%
  \institution{Clarkson University}
  \streetaddress{8 Clarkson Avenue}
  \city{Potsdam}
  \state{NY}
  \country{USA}
  \postcode{13699}
}

\renewcommand{\shortauthors}{Khan, Devlen, Manno, and Hou}

\begin{abstract}
 Utilization of the Internet in our everyday lives has made us vulnerable in terms of privacy and security of our data and systems.  Therefore, there is a pressing need to protect our data and systems by improving authentication mechanisms, which are expected to  be low cost, unobtrusive, and ideally ubiquitous in nature. Behavioral biometric modalities such as  mouse dynamics (mouse behaviors on a graphical user interface (GUI)) and widget interactions (another modality closely related to mouse dynamics that also considers the target (widget) of a GUI interaction, such as links, buttons, and combo-boxes) can bolster the security of existing authentication systems because of their ability to distinguish an individual based on their unique features. As a result, it can be difficult for an imposter to impersonate these behavioral biometrics, making them suitable for authentication. In this paper, we survey the literature on mouse dynamics and widget interactions dated from 1897 to 2023. We begin our survey with an account of the psychological perspectives on behavioral biometrics. We then analyze the literature along the following dimensions: tasks and experimental settings for data collection, taxonomy of raw attributes, feature extractions and mathematical definitions, publicly available datasets, algorithms (statistical, machine learning, and  deep learning), data fusion, performance, and limitations. Lastly, we end the paper with presenting challenges and promising research opportunities.
\end{abstract}



\maketitle

\section{Introduction}
Use of the Internet has become a part of our daily routine. Due to such intensity in internet usage, the risk of data breaches is higher than ever. An example of this is large-scale data breaches, such as those indicted at Yahoo and Equifax, that essentially exploited the weaknesses of the traditional username/password authentication scheme~\cite{the_united_states_department_of_justice_2017, the_united_states_department_of_justice_2020}. Therefore, it has become imperative for us to find additional ways to protect our data, which need to be low-cost, unobtrusive, and widely available. Physiological-based biometrics such as facial recognition, fingerprints, and iris authentication all require additional hardware and tools, which can be cumbersome due to cost and usability. However, mouse dynamics and widget interactions integrated modules are very inexpensive and unobtrusive technologies that can be implemented without interfering with day-to-day computer operations. Unlike existing knowledge-based authentication, such as passwords, that are based on \enquote{what you know}, these behavioral biometrics verify a user's identity based on \enquote{what you are}.

Biometrics, using human characteristics for identification and authentication purposes, has been a subject of scientific research for many years~\cite{jain2006biometrics}. There are two types of biometrics: one is based on physiological characteristics such as fingerprints, iris, facial recognition, and palm/hand geometry~\cite{adeoye2010survey}, and behavioral biometrics, such as mouse dynamics, widget interactions, keystroke dynamics, swipe dynamics, motion, and walking/gait~\cite{yampolskiy2008behavioural,khan2021authenticating}. The earliest example of behavioral authentication may be found in a 19\textsuperscript{th} century paper about telegraph typing, where a dispatcher of telegraph operations could identify operators based on the rhythmic sound of pressing telegraph keys~\cite{banerjee2012biometric,vacca2007biometric, bryan1897studies}. A biometrics system can be applied in two types of applications: 1) identification and 2) authentication. For identification, the system looks to find a one-to-many match by searching through all of the user templates to establish a person's identity. For authentication, a system verifies a person's identity with a one-to-one comparison between the current and existing template. In a nutshell, it verifies if the person is the same person who they claim to be~\cite{ross2006handbook}. Interestingly, Pusara and Broadley~\cite{pusara2007examination} categorize behavioral biometrics into further classifications: direct behavioral biometrics (e.g., mouse dynamics, GUI based events, keystrokes) and indirect behavioral biometrics (e.g., command line interface (CLI), call stacks).

In terms of behavioral biometrics, keystroke dynamics have been the center of research for a longer period of time relative to other behavioral biometrics. For example, Spillane from IBM during the 1970s performed a data collection process with respect to keystrokes for individual identification~\cite{spillane1975keyboard}. The following decades produced more survey papers regarding keystroke dynamics~\cite{teh2013survey, banerjee2012biometric} compared to mouse dynamics and widget interactions. In contrast, research about mouse dynamics only started to pick up by the beginning of the 21\textsuperscript{st} century and online interaction-based authentication has only been broadly researched since the beginning of the last decade~\cite{khan2021authenticating, hashia2005using, pusara2004user, pusara2007examination, imsand2008applications, revett2008survey, bours2009login, zheng2011efficient, zheng2016efficient, gamboa2003identity, mondal2013continuous, shen2009feature, shen2010hypo,shen2012continuous, shen2012effectiveness, shen2012user, shen2014performance, shen2017pattern, ma2016kind, kaixin2017user,almalki2019continuous, tan2017insights, tan2019adversarial, antal2019intrusion, antal2019user, antal2020mouse, antal2021sapimouse, salman2018using, hu2017deceive, hu2019insider, gao2020continuous, aksari2009active, ahmed2007new, ahmed2010mouse, sayed2009static, sayed2013biometric, chong2018mouse,everitt2003java, hinbarji2015dynamic, syukri1998user, pusara2007examination}. In our literature review, we have found only one brief mouse dynamics survey  by Revett et al.~\cite{revett2008survey}  that describes literature on mouse dynamics. In fact, Revett et al.~\cite{revett2008survey} only reviews four papers~\cite{pusara2004user,gamboa2003identity, ahmed2007new, hashia2005using} about users, raw data, feature extraction, algorithms and metrics. Therefore, a comprehensive up-to-date literature survey is justified for the research community. In addition, we also survey widget interactions as a new novel authentication modality.

This paper is organized as follows: Section~\ref{research} presents our paper collection methodology. In Section~\ref{A}, we  discuss mouse dynamics along with widget interactions as part of human psychology. In Section~\ref{data colelction}, we delve into data collection in terms of tasks and experimental settings as well as public datasets. In Section~\ref{raw data and feature}, we describe raw data and define different features mathematically. In Section~\ref{State_Based_Papers}, we survey the different statistical and pattern based algorithms on mouse dynamics as well as deep-learning algorithms. In Section~\ref{widget}, we survey widget interactions as a behavioral biometric. In Section~\ref{challenges}, we detail some of the challenges and research opportunities in mouse dynamics and widget interactions. Lastly, Section~\ref{conclusion} concludes our survey.

\section{Paper Collection Methodology}
\label{research}

We perform our literature survey by gathering all the pertinent papers via \textit{Google Scholar}. 
We surveyed a total of 123 papers directly for mouse dynamics and some additional papers related to psychology and HCI.
We follow a step-by-step process as follows. 
    We search \textit{Google Scholar} with the following  keywords and phrases (``mouse authentication'' AND ``survey'' (51 results), ``mouse dynamics'' AND ``signature authentication'' (50 results), ``mouse authentication'' AND ``psychology'' AND ``survey'' (6 results),  ``mouse dynamics'' AND ``intrusion detection'' (581 results), ``mouse dynamics'' AND ``computer security'' (573 results), ``mouse dynamics'' AND ``insider threat'' (183 results), ``mouse dynamics'' AND ``masquerade attack'' (33 results), ``mouse dynamics'' AND ``spoof attack'' (7 results)).
    Among these search results, we first select papers based on two conditions: papers that are of higher quality (with over 50 citations), yielding $\sim$114 papers, and papers that have less than 50 citations and are relatively new (since 2015), yielding $\sim$461 papers.
    The number of 50 citations was chosen based on our own experience; a paper published 8 years ago should have gathered more than 50 citations to be considered of ``good'' quality.
    We further select each paper based on their relevance to mouse dynamics and widget interactions and their overall quality, yielding a total of 123 papers. 
    We also review other relevant papers cited by the ones selected above.
    Lastly, we summarize this body of literature along the dimensions of data acquisition, feature extraction, datasets, algorithms and training/testing techniques, and performance measures.

\section{Psychology Behind Behavioral Biometrics: Mouse Dynamics and Online Behavior}
\label{A}
In this section, we survey the psychological impacts such as the cognitive, perceptual, and motoric understanding of humans as they generate mouse dynamics and widget interactions. 
\subsection{Psychology Behind Mouse Dynamics}

 As early as the 19$^{th}$ century, Bryan and Harter~\cite{bryan1897studies} first observed in their psychological experiments that a dispatcher of telegraph operations can identify a person based on the rhythmic sound of pressing keys. Subsequently, this topic based on interaction with a telegraphic device started to be picked up as a subject of experimental psychology, which aims to provide tangible validations of information theory using different measurement and modeling techniques, such as Fitts' and Hick's laws~\cite{hick1952rate, fitts1954information}. Since the 1970s, due to the advent of computers, experimental psychology has been used to reduce \enquote{Human-Computer Interaction} (HCI) time to save money and improve usability and operational efficiency. For example, Card, English and Burr~\cite{card1978evaluation} use mouse, joystick, step key and text key modalities to select a word  on a CRT (Cathode Ray Tube) by moving a cursor from a home position to a target position with varying width (W) and amplitude (A). They find that Fitts' law can be modeled after all four modalities; additionally, from their experiments, mouse is determined to be the fastest among all four modalities.

A substantial amount of HCI literature already exists, which tries to make user interactions more efficient. The goal of HCI is to improve efficiency of user interactions by optimizing the interaction speed (i.e., timing, often in terms of milliseconds). Though mouse dynamics is a behavioral biometric which also derives from user interactions with the computer, the goal is to authenticate users by investigating duration, as well as other features that interact with a system distinctly. Therefore, they both share some common metrics (i.e., interaction speed/timing), but for different purposes. This notion of interaction speed/timing has been studied in the field of experimental psychology called interaction ergonomics. For example, how quickly can a user position the mouse and click on a button or a drop-down menu. What would be the best way to minimize error rates during authentication for these events? For computer usage, interaction ergonomics tries to answer these aforementioned questions in the formulation of two \enquote{laws} (i.e., Hick's  and Fitts' laws) within  experimental psychology~\cite{revett2008survey}.

Hick's law stipulates that as the number of choices increases, so does the response time in decision making. Hick performed two types of experiments: the first which demonstrates if someone is well-trained at tasks; the response time should be largely proportional to the information extracted based on multiple choices, and the second, which is a ten-choice experiment, where a trained and an untrained user are convinced to reduce their reaction time by willfully making mistakes in selecting the given choices. In terms of a trained user, the response time is relatively constant with a smaller variance. Conversely, the untrained user has a longer response time with a larger variance. To execute the experiments, Hick provided a theoretical background of \enquote{quantity of information} based on Shannon’s information theory~\cite{shannon1949mathematical}, which is written as:
 \[H= - \sum_{1}^{i}P_i log P_i\]
\noindent where \textit{H} is the entropy or expected information, and \textit{$P_i$} is the probability of possible alternatives. The apparatus used in the experiments was ten lamps as choices in an irregular circle and Morse Keys to provide responses via pressing~\cite{hick1952rate}. Hick's law has been utilized into current experimental psychology when designing HCI regarding multiple choices. Throughout Hick's experiments, he tried to prove with some degree of evidence that the response time is proportional to the extracted information. However, the relationship between response time and the mode of operation is still unclear. For example, for a trained operator, the response time is proportional to the extracted information due to familiarity with the system (i.e., mode of operation); however, for an untrained operator, the response time may not be proportional to the extracted information due to their unfamiliarity with the system~\cite{salthouse1986perceptual}.

 Fitts initiated multiple experiments involving subjects to make successive responses based on a particular movement that encompass the speed/accuracy tradeoff. In the first experiment, the subject had two stylus of different weights touching two plates with varying widths as targets, which were spread out over varying distances. In the second experiment the subject placed a disc over a pin from one position to another. In the third experiment, the subject transferred the pin from one hole to another. To measure the difficulty level of performing a task and performance rate in all three experiments, an  index of difficulty and performance rate were used, which are defined below:
 \[I_d = log_2\frac{2A}{W_s}\hspace{0.1cm}bits/response, I_p = \frac{1}{t}log_2\frac{2A}{W_s}\hspace{0.1cm}bits/sec\]

\noindent where $I_d$ is the level of difficulty, $W_s$ width of the target,  \textit{A} distance from the pointing device to the target, $I_p$  performance rate, and \textit{t}  average duration. 
Fitts observed that as the distances increase and the widths are kept constant for the target or vice versa, $I_d$ of performing such a task increases. Also, $I_p$increases to a certain level in the beginning, but then falls out of range. This observation is the same across all three experiments. Fitts also demonstrated in all the experiments that the length of time in moving an object between targets increases as the distance between them increases. These observations categorize the foundation of the speed-accuracy tradeoff, which describes that as movement times get shorter and as targets size decreases; error rate increases. In all the experiments, the subjects performed in high performance scenarios with a timer. Fitts’ law has been applied into current experimental psychology when using a computing pointer device such as a mouse to make HCI more efficient~\cite{fitts1954information, revett2008survey, card1978evaluation}. However, it is unclear at what point the performance rate of Fitts' law will drop and how that will impact the mouse operations in a real-life scenario.

Apart from Fitts' and Hick's laws, which focus on predicting movement time, movement accuracy and uncertainty (commonly associated with pointing/steering tasks) are other important aspects of psychological and behavioral research in HCI. Many studies focus on static target tasks, where the target does not move. These tasks are similar in nature to many of the everyday tasks people perform on their computers (i.e., clicking a desktop icon or button). As aforementioned, Fitts' law encapsulates the idea of the speed-accuracy tradeoff, which is one of the most important aspects of human performance. MacKenzie \cite{mackenzie1992} and Zhai et al. \cite{zhai2004} examined the potential of using statistical properties of pointing accuracy to improve the performance of Fitts' law. MacKenzie emphasized a previously underutilized model that employs a normalization of target width, known as effective width $W_e$, based on the subject's actual behavior (output condition), rather than expected behavior (input condition). Zhai et al. aimed to form a more complete model that encompassed two different layers of the speed-accuracy tradeoff. The first task-specific layer pertained to the accuracy requirements imposed by the task itself, while the second subjective layer is representative of individual biases in precision that are independent of the task requirements. Zhai et al. also introduced the idea of target utilization, which describes how the spread of movement endpoints deviates from target width. Wobbrock et al. \cite{wobbrock2008} focused on modeling error rate based on pointing accuracy (i.e., whether a button was hit or missed) as derived from Fitts' law and with use of effective width. Zhou et al. \cite{zhou2009} discussed movement accuracy with tasks other than pointing as well. Zhou et al. focused on temporally constrained (time constrained) trajectory based tasks (tracing a line with a touch stylus on a touchscreen along tunnels of various widths and lengths). Lee et al. \cite{lee2016, lee2018} studied temporal pointing, a type of pointing that requires minimal or no spatial aiming, imposes a time constraint for the task, and where input is a discrete event such as a button press. In these studies, they focused on modeling movement uncertainty in dynamic target tasks. Zheng found the mean time of a user's actions (a sequence of point-and-clicks over textual URLs) was proportional to the index of difficulty, which followed Fitts' law in a real world environment. Zheng \cite{zheng2014exploiting} empirically validated Fitts' law by collecting data from users on a web forum in an unrestricted manner using a mouse to point and click as an action on text-based links. Zheng's work is the only one that validated Fitts' law for mouse authentication datasets, but it was not used directly as a feature in mouse authentication.  

The principles of these works have the potential to be applied to mouse dynamics authentication. Considering the width of a target a user is trying to click as well as the amplitude of distance from the start position of the cursor to the target could be useful features following the ideas of $W$ and $A$ in Fitts' law. The principle of effective width $W_e$ as discussed by MacKenzie and Zhai et al. could similarly be used as a feature for classification in mouse dynamics. Effective width shows potential for better characterizing a users behavior since it is based on the measured behavior of a user rather than their expected behavior. Moreover, Zhai et al. showed $W_e$ encompassed both layers of the speed-accuracy tradeoff better than $W$ with different levels of target utilization. The principle of user error prediction as described by Wobbrock et al. could also be adapted to mouse dynamics. Examining the distribution of movement endpoints and determining frequency of error and by what margin could serve as another feature for classification. Another possible feature that could be derived is variability of accuracy, defined as predicted error versus actual. Zhou et al. didn't focus on traditional pointing tasks, but the idea of examining an entire mouse trajectory and again looking at amplitude of distance of the trajectory as well as width of trajectory rather than the width of target could also serve as a feature for classification. Lee et al. didn't focus on tasks where typical features such as velocity and acceleration could be determined, but systems utilizing a fixed cursor like some games do could potentially take advantage of the presented error model in order to better design targets. This area of research is yet to be widely explored, with only Zheng examining Fitts' law as it deals with mouse cursor applications, and as such the greater research community would benefit from more studies in this area.

Another area of research studying the speed-accuracy tradeoff deals with dynamic target based tasks. These tasks involve targets that are not fixed in one position on the screen, and instead are in motion. Jagacinski et al. \cite{jagacinski1980} and Hoffmann \cite{hoffmann1991} both focused on estimating movement time with dynamic target tasks. Jagacinski et al. focused on dynamic tasks with temporal constraints, which are classified by use of a moving target and the need to capture that target within a set time period. Jagacinski et al. also created a new model for index of difficulty $I_d$ incorporating components such as movement time $MT$, distance amplitude $A$, width of target $W$, velocity of target $V$ and three regression coefficients. This model better fit dynamic target tasks than Fitts' original model. Subsequently, Hoffmann mathematically derived two models for predicting movement time with an attempt to take into account the effect of steady state position error on effective target width. These models had shared components of variable of gain $K$, velocity $V$, and effective width $W_e$, with the second model containing additional empirical constants. By applying these models to Jacaginski et al.'s data, Hoffmann found that there was a significantly lower rate of information processing during the accuracy phase compared to the distance-covering phase of the motion. Huang et al. conducted a series of studies to model movement uncertainty and predict error rates in various target selection tasks. In their first study \cite{huang2018}, they introduced a Ternary-Gaussian model for 1-D unidirectional moving targets, representing the endpoint distribution as a sum of three Gaussian components: one reflecting user bias, another accounting for uncertainty due to target movement, and the third related to task precision and movement speed. In their subsequent study \cite{huang2019}, the researchers extended their modeling to 2-D moving target selection using a 2-D Ternary-Gaussian model, which accounted for endpoints in a new ``velocity coordinate system''. Finally, in their investigation of crossing-based moving target selection \cite{huang2020}, they proposed a Quaternary-Gaussian model, similar to the 1-D Ternary-Gaussian model but adapted for crossing-based selection. Crossing-based selection is selection of a target based on crossing through it, instead of using a discrete means such as pressing a mouse button. Park et al. \cite{park2020} presented an Intermittent Click Planning (ICP) model that described the process by which users plan and execute click actions and could be directly used for predicting error rate. This model had components of target velocity $v_t$, target position $p_t$, target width $W$, pointer velocity $v_p$, pointer position $p_p$, and mean period of click repetition $P$. This model was also successfully employed to discern cognitive differences among players of a first person shooter game. It was shown that in users playing PlayerUnknown's Battlegrounds that gamers and non-gamers had a similar level of ability to estimate click timing from visual cues, but gamers have better ability to encode the rhythm of clicks with an internal clock. 

Dynamic tasks are not yet highly prevalent in the realm of mouse dynamics authentication. However, dynamic authentication systems as a one-time authentication method could be contrived for which the models presented for dynamic targets have  potential use. The models for estimating movement time proposed by Jagacinski et al. and Hoffmann et al. have components that could serve as features for classification. Again, the use of distance amplitude $A$ and width $W$ could serve as useful features for classification in the context of the movement trajectory, as well as velocity $V$ of the moving target. The variable of gain $K$ in regards to steady state position could also potentially serve as a useful feature, and effective width $W_e$ also was shown to be of good performance for modeling these tasks. The Gaussian models that Huang et al. proposed could  be of use for considering the individual endpoint distributions of dynamic target acquisition tasks in 1D and 2D, which could reveal a user's tendency to hit or miss targets of certain widths and of certain distances from starting points. The models Lee et al. focus on are not of a nature typically found in mouse dynamic tasks, as there is no cursor movement and as such no features to be derived from movement, but the principles of selection uncertainty studied could have potential use for deriving features for mouse dynamics authentication. The ICP model Park et al. proposed also has components that could be used as features for dynamic target tasks, and shows high potential due to already having been used to identify cognitive differences in users. From these studies, it is evident that HCI principles and elements of human psychology have great potential to help classify individuals in a mouse dynamics authentication system, across multiple dimensions, different selection methods, and with both dynamic and static targets.

\subsection{Psychological Modeling of Online Behavior}

Similarly, widget interactions evolve from HCI. Widget interactions are a special kind of interaction such as hovering over via a cursor on certain widgets (e.g., a button or icon) as a part of online behavior. As mentioned earlier, HCI has been studied for the purpose of improving operational efficiency. On the other hand, the goal of widget interactions is to authenticate users during online activities utilizing time (i.e., milliseconds) as features. Both of them also have common metrics (i.e., timing/duration) like mouse dynamics. There are different kinds of online interactions, such as pageview sessions, monitoring online activities of users (to increase sale), and structured agnostic interaction. Structured agnostic interaction is when a person browses randomly via a mouse on a website for the purpose of extracting mouse movement, speed, curvature, trajectory, etc., instead of widget interactions. In this context, widget interaction behavior is often ignored, and that makes it a novel authentication method~\cite{khan2021authenticating, imsand2008applications}.

People use different modeling techniques to improve the efficiency of their operations. Efficiency and authentication are inter-related due to the measurement techniques that can be used. One such technique is the GOMS (Goal, Operator, Method and Selection) theory, which is to predict and explain real world experiences of HCI quantitatively through the lens of experimental psychology. By definition, a goal is considered a well-known task within the model, an operator is who performs the task, a method is a way to perform the task, and a selection is to have multiple options to reach the final goal. Due to GOMS' ability to quantify human performance effectively in earlier years, it has become a pioneer to improve the efficiency of HCI by eliminating unnecessary user actions~\cite{olson1995growth}. There are several variants of GOMS, but among the major ones are: CPM-GOMS (Critical Path Method), KLM-GOMS (Keystroke-Level Model) and NGOMSL (Natural GOMS language). CPM-GOMS extends the GOMS method to model parallel activities among perception, cognition and motor functionalities~\cite{gray1993project}. KLM-GOMS is considered the simplest  variant of GOMS. It uses basic operators like keystrokes, cursor movement time, button presses, and double clicks to estimate task times~\cite{john1996goms}. Lastly, NGOMSL depends on execution times that are collected based on software by a user performing the tasks with the system~\cite{wu2008queuing}.

The GOMS theory considers routine perceptual, cognitive and motoric operations of a user for its analysis, such as routine timing information of a keystroke, a single mental operator (i.e., time) to extract the next piece of information from memory, or pointing to a target using a mouse on a display~\cite{olson1995growth}. Gray and Boehm-Davis~\cite{gray2000milliseconds} demonstrates that HCI behavior comprises interaction between artifacts (e.g., a mouse and button) and elementary human cognitive, perceptual, and motoric operations. Their research shows that interactive behavior can be optimized with alternative techniques, such as moving the cursor of a mouse or clicking a button in a certain way, to reduce the cost or time of that interaction and save milliseconds to make HCI more efficient. However, there are shortfalls of using GOMS, such as the lack of parallel modeling based on human perception, cognition and motoric activities; therefore, they use CPM-GOMS to model the parallel activities effectively to reduce time.

Gary, John and Atwood~\cite{gray1993project} experiments with the CPM-GOMS model on toll and assistance operators (TAO) to find out the possibility of saving \$3 million for a company. The major motivation behind the proposal is to optimize interaction with the proposed workstations by reducing average work time per call, which would offset the capital cost compared to older workstations. They use CPM-GOMS to model parallel behaviors displayed by TAOs. Using CPM-GOMS, they are able to determine the proposed workstations would cost \$2 million more than older workstations due to the difference in keyboards and screen layouts. As a result, the model is able to show that learning very small differences within the system design can help minimize the cost of operations.

Overall, the purpose of GOMS theories is to improve the efficiency of HCI and the usability of a system. However, as one can see while experimenting with different artifacts/widgets (e.g., different buttons) on a GUI, this provides a conceptual understanding of widget interactions. Hence, we believe that these theories for HCI have provided the very foundation of the widget interactions modality to this day.

\section{Data Collection for Mouse Dynamics: Nature of Tasks and Experimental Settings}
\label{data colelction}

Data collection is an inherent part of mouse dynamics research. Many factors can be involved during data acquisition processes. In this section, we categorize the techniques that researchers have applied in collecting raw mouse data, in terms of the nature of the tasks that users are required to perform, as well as the experimental settings~\cite{ross2006handbook}. 

\subsection{Nature of Mouse Data Collection Tasks}

The nature of mouse data collection tasks may vary (e.g., different tasks, different apps), which can influence user behavior during  data collection~\cite{ahmed2007new, feher2012user, antal2019user, almalki2019continuous, pusara2004user}. Therefore, we divide methods of data collection into four categories based on user actions and the app/data collection software involved as follows: fixed static sequence of actions, app restricted continuous, app agnostic semi-controlled and completely free data collection. 
\begin{enumerate}
    \item  \textit{Fixed static sequence of actions} represents fixed and repeated mouse actions based on the predefined tasks setup by the researchers via apps (i.e., moving a mouse cursor from fixedly situated button A to button B presented by the experiment). For instance, in Shen et al.~\cite{shen2014performance, shen2012effectiveness, shen2012user} and Ancien et al.~\cite{acien2020becaptcha}, all the participants are asked to perform a fixed set of mouse operating actions to produce patterns. These fixed set of actions are made of 8 consecutive movements isolated by single and double clicks. As a result, the actions generate directions in which each one of them is 45-degrees out of a 360-degree range on a screen (see Figure~\ref{static authentication_2}). 
    Other studies that fall under this category are \cite{carneiro2015using, sayed2009static, sayed2013biometric, kaminsky2008identifying, revett2008survey, aksari2009active, hashia2005using, syukri1998user, antal2021sapimouse}. 

\begin{figure*}
\centering
     \begin{subfigure}[b]{0.45\textwidth}
         \centering
         \includegraphics[width=\textwidth]{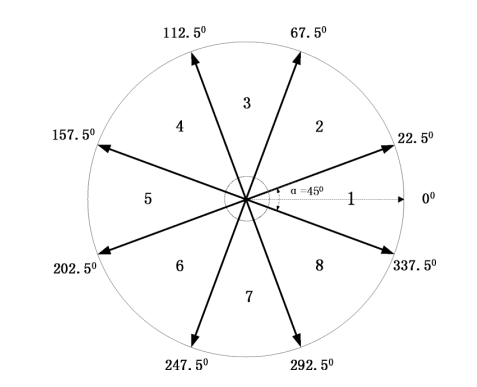}
         \caption{Mouse Operation Mode Based on Static Mouse Operating Patterns~\cite{shen2012user}}
         \label{static authentication_2}
     \end{subfigure}
     \hfill
     \begin{subfigure}[b]{0.4\textwidth}
         \centering
         \includegraphics[width=\textwidth]{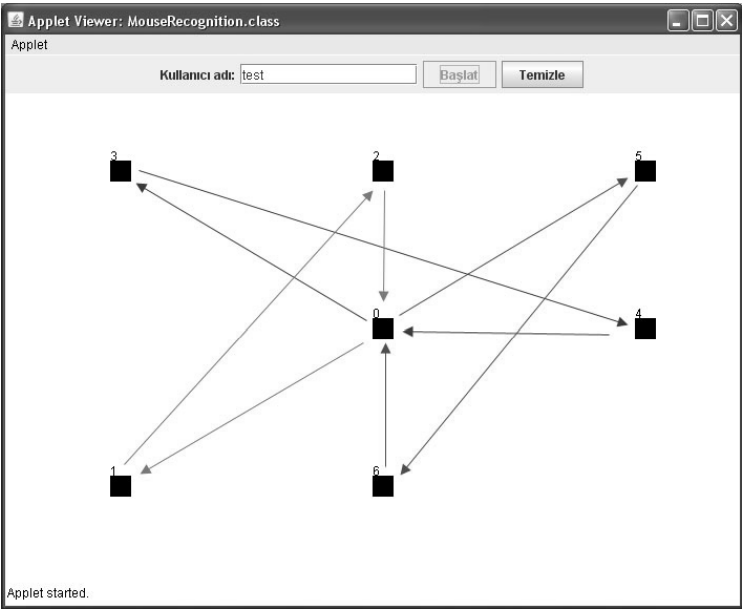}
         \caption{Paths Between Squares that a User Must Follow~\cite{aksari2009active}}
         \label{static authentication_aksari}
     \end{subfigure}
     
     \begin{subfigure}[b]{0.6\textwidth}
         \centering
         \includegraphics[width=\textwidth]{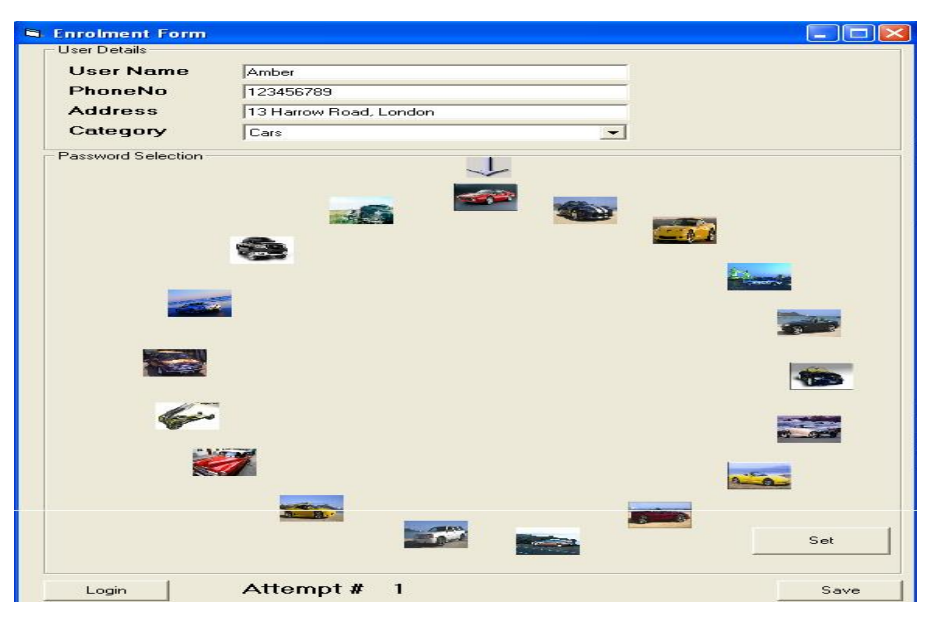}
         \caption{Static Authentication of a Mouse-Lock using Thumbnail Images of Cars~\cite{revett2008survey}}
         \label{static authentication_1}
     \end{subfigure}
\caption{Examples of Fixed Static Sequence of Actions}

\end{figure*}

\begin{figure*}
\centering
     \begin{subfigure}[b]{0.5\textwidth}
         \centering
         \includegraphics[width=\textwidth]{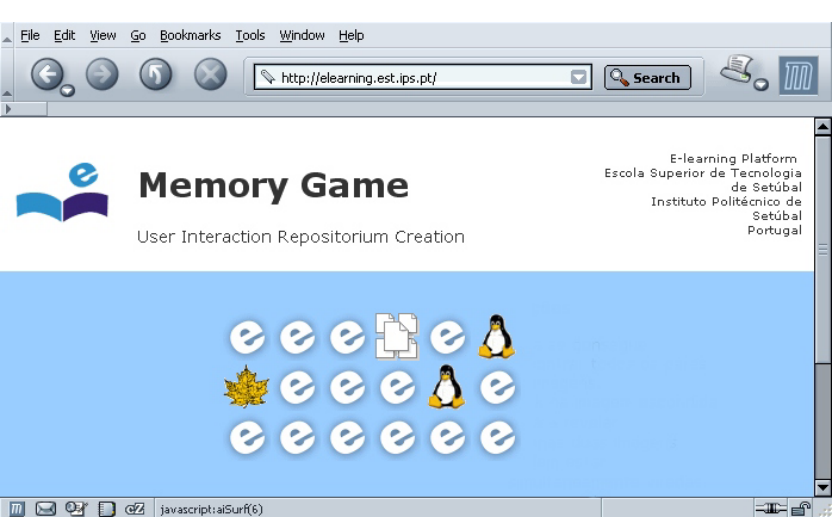}
         \caption{A Memory Game Based on Static Authentication~\cite{gamboa2003identity}}
         \label{static authentication_3}
     \end{subfigure}
     \hfill
     \begin{subfigure}[b]{0.45\textwidth}
         \centering
         \includegraphics[width=\textwidth]{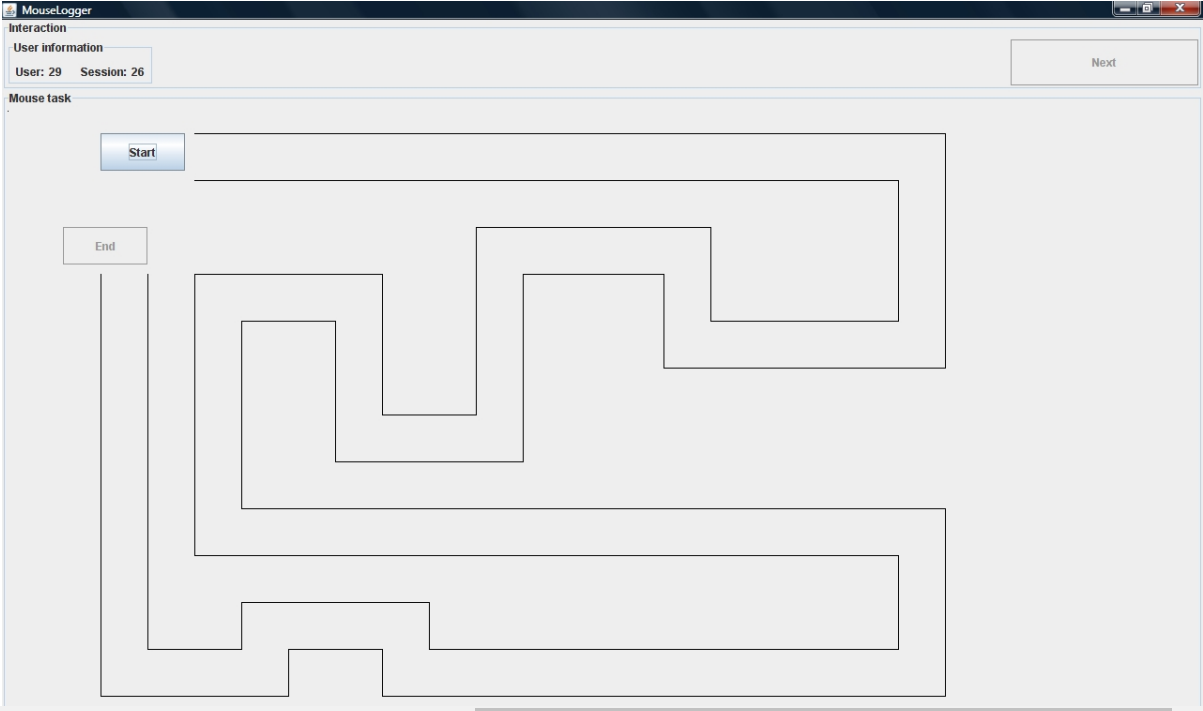}
         \caption{Navigating the Mouse Pointer following GUI Based Maze~\cite{bours2009login}}
         \label{dynamic authentication}
     \end{subfigure}
     \caption{Examples of App-Restricted Continuous Authentication}

\end{figure*}
    \item \textit{App restricted continuous} data collection represents the scenario where collection  limited by the conditions of the specific app. For example, according to Gamboa and Fred~\cite{gamboa2003identity}, they develop a game of grid of 3x6 tiles (i.e., condition of the game), where a user has to match a pair of tiles by clicking on them (see Figure~\ref{static authentication_3}). Other studies that fall under this category are \cite{jorgensen2011mouse, bours2009login, ma2016kind, pusara2004user, chong2019user, chong2018mouse, kaminsky2008identifying, tan2019adversarial}.
    
    \item \textit{App agnostic semi-controlled} data collection represents the type of task that does not depend on the app, per se. However, it depends on many factors other than the app, making the data collection process semi-controlled. For example, Shen et al.~\cite{shen2009feature} describe many factors that contribute to the variability of mouse dynamics data collection aside from apps, such as different brands of mice, GUI settings, emotional state (e.g., anger, despair, happiness, stress, relax), physical condition (e.g., tiredness, illness), distance between the mouse and body, height of the chair and screen size of the computer. Other studies that fall under this category include \cite{shen2012continuous, ernsberger2017web, kaixin2017user, zheng2011efficient, zheng2016efficient, antal2019user}

    \item \textit{Completely free} data collection represents collection that is free of any set of guidelines for data collection process. For example, Ahmed and Traore~\cite{ahmed2007new, ahmed2010mouse} collect data using their own software from participants without any restriction. They ask the users to install the software on their computers and perform their routines, but varying activities (e.g., web browsing, word processing). The software runs on the background and monitors user's activity and send to a remote server unobtrusively. Other studies that fall under this category include \cite{salman2018using, mondal2013continuous, shen2017pattern, shen2010hypo, zheng2011efficient, zheng2016efficient, hu2017deceive, tan2017insights, almalki2019continuous, antal2019intrusion, gao2020continuous, hu2019insider, chong2019user, chong2018mouse, antal2020mouse, nakkabi2010improving, schulz2006mouse, everitt2003java, hinbarji2015dynamic, antal2020mouse}
    \end{enumerate}
    From the above classifications, although we can delineate that it may be beneficial in some cases to have a fixed static sequence of actions or app restricted continuous data collection to reduce noise or remove outliers from the experiments, these collection types may not be the best choices due to lessening or restricting the authenticity of a user's behavior. From the empirical perspective, conducting experiments should be geared towards more realistic approaches, such as app agnostic or completely free data collection, to represent the authentic behavior of a user.

\begin{figure*}[b]
\centering
   \includegraphics[width=4.5in]{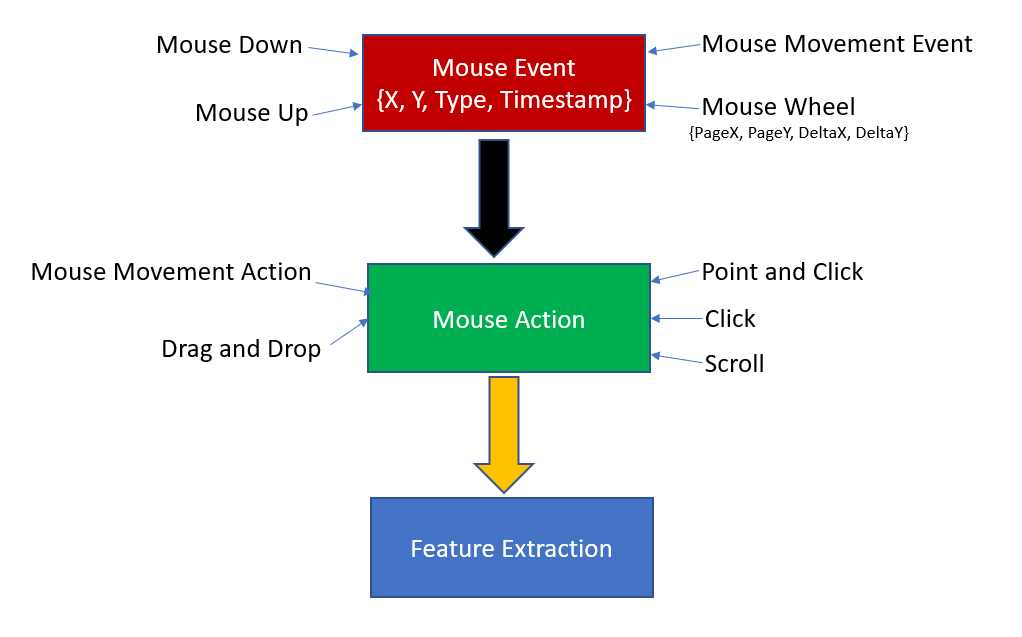}  
   \caption{Raw Data Processing Pipeline: from Mouse Events and Actions to Features}
   \label{Raw Attribute}

\end{figure*}
\subsection{Experimental Settings}

Experimental settings have been one of the key parts for a proper mouse authentication system. Researchers refer to different brands of mouses, different sizes of screen layouts, variations of different movement rhythms of a mouse by the user, temperature and lighting of the room, as well as different types of computers and human emotions as parts of experimental settings~\cite{almalki2019continuous, ahmed2007new, ma2016kind, belman2019insights, bours2009login, mondal2013continuous}. An experimental setting is divided into three categories: uncontrolled~\cite{bours2009login}, semi-controlled~\cite{khan2021authenticating}, and controlled~\cite{shen2014performance}. The uncontrolled setting can be defined as an environment where nothing is controlled in the way a user provides information.  For instance, a user is asked to download the software on his/her computer (laptop/desktop) and perform routine tasks freely without any observation from the researcher or any other internal/external influence (e.g., room, temperature, chair, computer, lighting). Uncontrolled setting provides the most realistic scenario for tested and deployed systems~\cite{bours2009login,zheng2011efficient, zheng2016efficient}. Semi-controlled setting is a little different than controlled and uncontrolled settings. It is where some of the aspects of the experiment are controlled and some are not. For example, participants are remotely logged into a computer to perform the tasks, where the usage of the software is the same, but the surrounding environment for each participant is different~\cite{khan2021authenticating}. Controlled setting is where the participants are asked to partake in the same environmental setting by the researchers. Same environmental setting means same lighting setup of the room, same computer with software installed to record mouse actions, same height of the chair, more or less same distance from the mouse to the body as well as same controlled temperature of the room for all the participants~\cite{shen2012user, revett2008survey, shen2009feature}. 

\section{Raw Data, Features, and Public Datasets}

\label{raw data and feature}
After data collection, researchers conduct further analysis of the collected raw data to extract features, to create substantial meaning for the classifiers to comprehend and perform binary authentication. In this section, we expound classifications of raw data into different categories and how we perform feature extractions using raw data.

\subsection{Raw data: Mouse Events and Actions}

\begin{figure*}
\centering

   \includegraphics[width=5in]{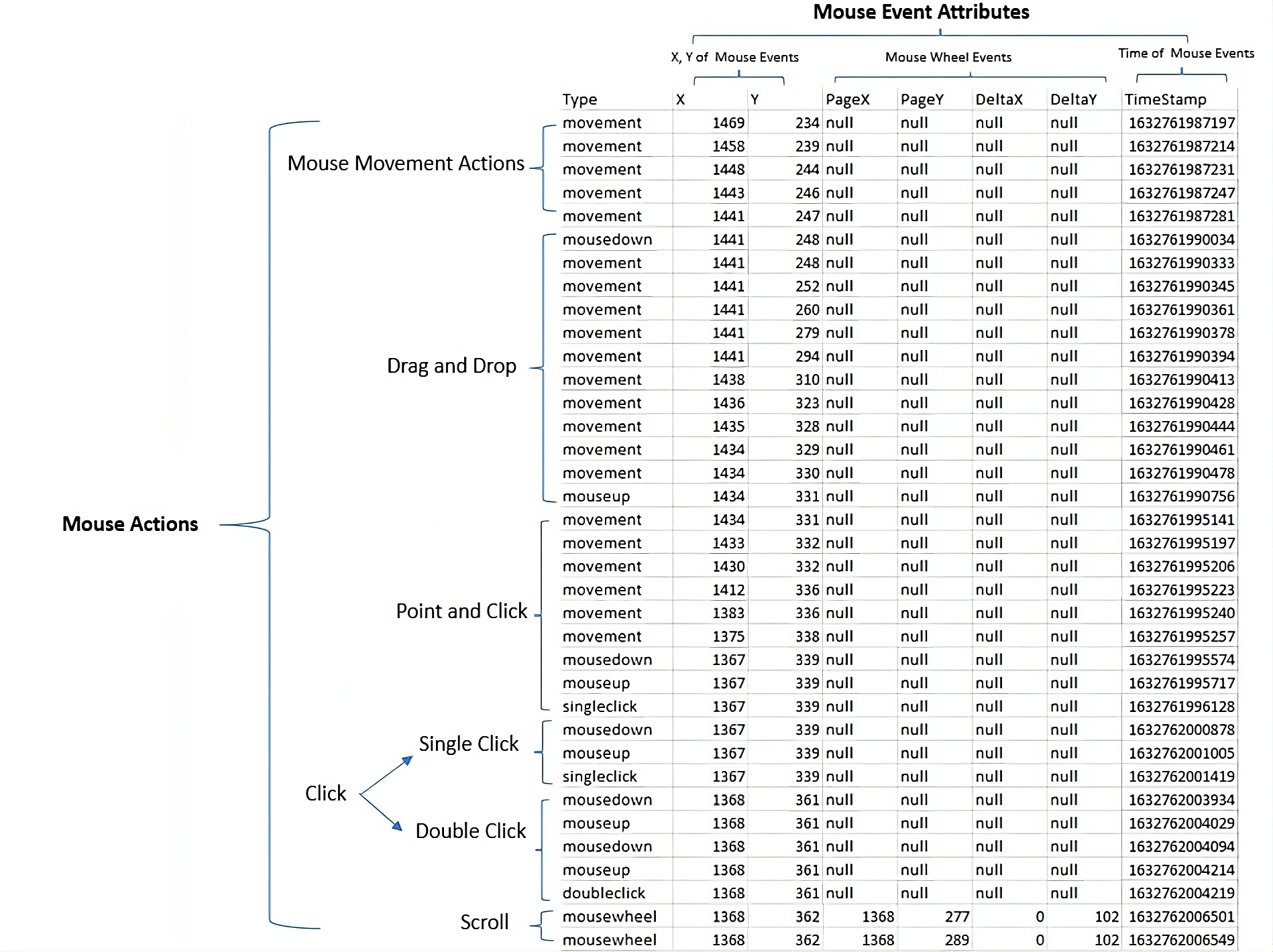}

   \caption{Samples of Raw Data From a Real User: Mouse Events (e.g., type, x, y, pageX, pageY, deltaX, deltaY and timestamp) on the top and Actions on the left (e.g., MM, DD, PC, C and S)}
   \label{Raw Data}

\end{figure*}

Figure~\ref{Raw Attribute} illustrate two categories of low-level raw data (i.e., mouse events and mouse actions). The mouse event is the lowest level as component raw data, which can be grouped together into meaningful actions as the next level of raw data, which is mouse actions.
Mouse raw data is usually captured by a software that runs in the background within a computer.  This software is normally custom-made based on the researcher's specifications, which is used to generate sequences of mouse event data, such as mouse down/up, mouse movement and mouse wheel by the user. These data are analyzed as system messages that define the event type, location and time of the mouse cursor. Practically, these generated system messages or low level raw data are not useful enough for analyzing human behavior for classifications. Therefore, they are usually grouped or combined into the next low-level raw data, called mouse actions, to extract features~\cite{shen2012continuous}. These mouse actions can be defined as :
\begin{figure*}[ht]
\centering

   \includegraphics[width=4in]{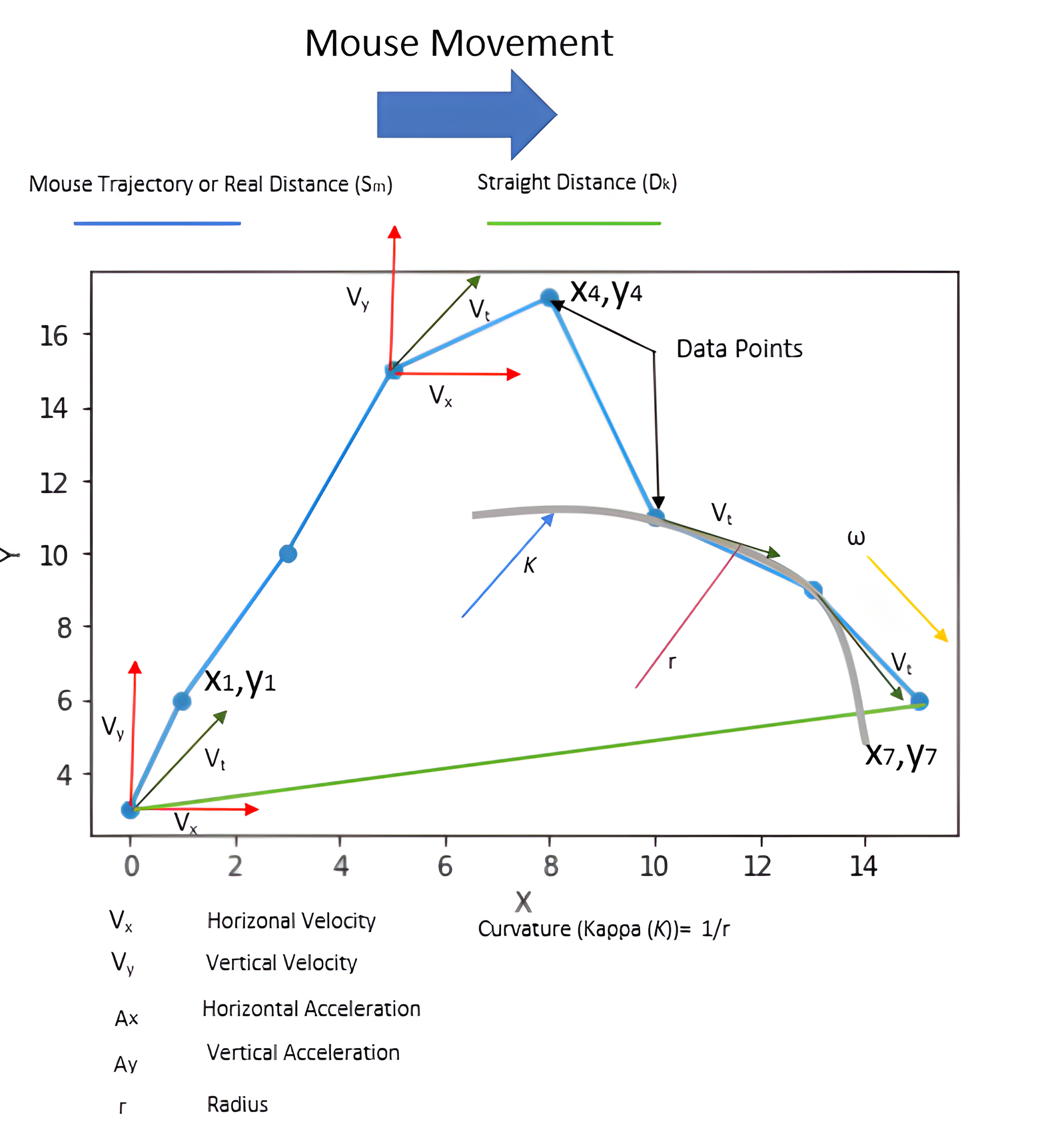}

   \caption{The Mouse Movement (MM) Consists of Data Points That Define a Trajectory}
   \label{mouse_1}

\end{figure*}
\begin{itemize}
    \item \textit{Mouse Movement} (MM) actions: 
    The mouse movement can be defined as location changes of a mouse without pressing any mouse up or down button~\cite{ahmed2007new, antal2019user, mondal2013continuous}. They can also be defined as a mouse trajectory or data points of a time-series~\cite{mondal2017study}. It is normally up to the researcher to decide how many coordinates would consist of a movement or trajectory (i.e., selecting a certain window size of coordinates or a timing condition).
    \item \textit{Drag and Drop} (DD): 
    The drag-and-drop is defined as starting the mouse with the button pressed, then dragging it to generate some movement until releasing the button. In other words, it is a sequence of mouse button down, movement and mouse button up~\cite{ahmed2007new, antal2019user, mondal2013continuous}; it can also be called a stroke~\cite{gamboa2003identity} or mouse move DD~\cite{feher2012user}.
    \item \textit{Point and Click} (PC): 
    When a user points or moves and clicks on an icon or a menu, he/she clicks by pressing and releasing the mouse button, which is called a point-and-click (see Figure~\ref{Raw Data}). PC can be subdivided mainly in two categories: point and single click, and point and double click. They can be further subdivided into point/right click or right double-click and point/left click or left double-click~\cite{feher2012user}.
    \item \textit{Click} (C): 
    This raw attribute demonstrates only a click without any movement before or within it. It can also be called a pause and click (i.e., where a user pauses for a certain time before using a single/double click or duration between mouse up/down)~\cite{hu2017deceive,zheng2011efficient}. Researchers also have been using time to click, paused time, the number of pauses and paused ratio (ratio between the number of pauses and the total duration of the click) before clicking as features~\cite{gamboa2003identity, zheng2011efficient} for feature extractions.
    \item \textit{Scroll} (S): 
    Scroll is when a user rolls a wheel on top middle of the mouse to move up or down on the web page. In other words, a scroll consists of a sequence of mouse wheels. For example, Figure~\ref{Raw Data} shows  PageX and PageY to provide coordinates of the mouse wheeling in x and y directions. Additionally, delta values indicate if the direction of the mouse cursor with respect to the web page is up or down. The delta value can be negative or positive based on a user's mouse wheeling activity~\cite{shen2012continuous, pusara2004user,detla}.

\end{itemize}

  Figure~\ref{Raw Data} depicts sample mouse data, from which one can observe a sequence of mouse events; and events are grouped into actions. Furthermore, it is evident that every action stems out of mouse movement except clicks, which is measured by duration ~\cite{antal2019user}. 



















\subsection{Features}

Figure~\ref{mouse_1} provides a visualization of a typical mouse trajectory, which demonstrates MM actions (Mouse Movement) along with discrete data points for features to be calculated. MM is the most dominant raw attribute to be included in most of the extracted features (although DD, PC, S are different actions, they are all to be extracted for features because of MM). C is the only that can be extracted based on duration without any MM~\cite{antal2019intrusion, antal2019user, ahmed2007new, gamboa2003identity, shen2009feature}. Table~\ref{feature definition} surveys features related to mouse trajectory as a list of feature names, definitions and mathematical formulations. Moreover, it is also common to use standard statistics such as minimum (min), maximum (max), median , mean, max-min, standard deviation (std), variance, skewness and kurtosis as features as well~\cite{ernsberger2017web, gamboa2003identity, feher2012user}.  In the following, as an example  we illustrate how to extract one feature for curvature. 

Curvature ($\kappa$): Geometrically, curvature can be described as one over radius $r$ as if moving along a perfect circle.  It can also be described as trajectory curvature~\cite{feher2012user}. In Figure~\ref{mouse_1}, we can approximate the curvature as a small arc of a circle. The radius of the curvature changes as we move along a trajectory based on different data points. The curvature will be zero if a curve is a straight line and radius will be $\infty$. In discrete terms, curvature formula is given as~\cite{math24_2021},
    $$\kappa = \frac{x'_{(2)}y''_{(3)} - y'_{(2)}x''_{(3)}}{({x'^2_{(2)}} + {y'^2_{(2)}})^{3/2}}$$ Taking the absolute value of $\kappa$ will provide the size of the curvature and its radius can be measured as,
       $$r = 1/|\kappa|$$
    As a result, we can literally calculate the curvature using features $x'_{(2)}, y'_{(2)},x''_{(3)}, y''_{(3)}$ as formulas from Table~\ref{feature definition} by measuring specific data points along the trajectory and plugging them into the curvature equation.

\renewcommand{\arraystretch}{0.5}

\begin{table}

\caption{Mouse Movement (MM) Features Extracted from Raw Data}

\label{feature definition}

\begin{tabular}{|p{40mm}|p{55mm}|p{55mm}|}

\hline
\small{\textbf{Feature Name}} & \small{\textbf{Feature Definition}} & \small{\textbf {Mathematical Formulation}} \\
\hline\hline
Traveled Distance ($D_{i}$) &  Distance between two adjacent mouse positions~\cite{shen2012effectiveness, antal2019user} & $\sqrt{ (y_{i+1}- y_{i})^2 +  (x_{i+1}- x_{i})^2}$\\\hline
Curve length/Real Dist. (${S_{n}})$& Length of  trajectory traveled~\cite{mondal2017study, carneiro2015using} & $\sum_{i=0}^{n} D_i$ \\\hline
 Elapsed Time/Mouse Digraph &  Time elapsed between two events~\cite{feher2012user, antal2019user,zheng2016efficient,carneiro2015using, revett2008survey} & $\sum_{i=0}^{n-1} (t_{i+1}- t_{i})$\\\hline
Movement Offset & Difference between distance along the curve $S_{i}$ and straight distance $D_{i}$~\cite{shen2012user, antal2019user} & $ S_{i}-D_{i}$\\\hline

Deviation Distance &Maximum distance from  mouse trajectory to  straight line~\cite{ma2016kind}& $\frac{(y_{0}-y_{1})x+ (x_{0}-x_{1})y+(x_{0}y_{1}-x_{1}y_{0})}{\sqrt{(x_{1}- x_{0})^2 +  (y_{1}-y_{0})^2}}$\\\hline
Straightness, Efficiency  & The ratio of  straight line distance to trajectory distance~\cite{gamboa2003identity,hinbarji2015dynamic}& $ \frac{D}{S_{n}}$\\\hline
Jitter & Tremor of  movement  measured as the ratio of the interpolated path length $I_{n}$ and  trajectory path $S_{n}$~\cite{gamboa2003identity, feher2012user} &   $ \frac{I_{n}}{S_{n}}$\\\hline
Velocity ($V$) &  Magnitude and direction of how fast a mouse cursor is moving~\cite{shen2012effectiveness,Differentiation, calculus_book, Trajectory_Calculation} & $\frac{D_{i}} {t}$\\\hline
Horizontal Velocity ($x'_{(2)}$)&  Movement speed in x~\cite{shen2012effectiveness, feher2012user,calculus_book, Trajectory_Calculation} & $ \frac{x_{2}-x_{1}} {t_{2}-t_{1}}$\\\hline
Vertical Velocity ($y'_{(2)}$) &  Movement speed in y direction~\cite{shen2012effectiveness, feher2012user} & $\frac{y_{2}-y_{1}} {t_{2}-t_{1}}$\\\hline
Horizontal Acceleration ($x''_{(3)}$)&  Acceleration in x direction~\cite{shen2012effectiveness, feher2012user,calculus_book, Trajectory_Calculation} & $ \frac{x_{3}-2x_{2} +x_1} {(t_{3}-t_{2})^2}$\\\hline
Vertical Acceleration ($y''_{(3)}$) &  Acceleration in y direction~\cite{shen2012effectiveness, feher2012user,calculus_book, Trajectory_Calculation} & $\frac{y_{3}-2y_{2} +y_1} {(t_{3}-t_{2})^2}$\\\hline

Speed against distance & Speed as function of traveled distance; obtained through periodic sampling using histogram and represented using discrete form of $Speed_i$ over distance~\cite{ahmed2007new, shen2012effectiveness}& $(Speed_i, S_i), i= 0,...n$ \\\hline
Average speed against distance & Average speed as function of  traveled distance
~\cite{ahmed2007new, shen2012effectiveness, feher2012user}& $(Speed_{{(0,i)}_{avg}}, S_{i}), i=0,...n$ \\\hline

X Acceleration against Distance &   Acceleration relative to traveled distance in x direction; can be obtained through periodic sampling and represented in discrete form of acceleration with distance~\cite{shen2012effectiveness}& $ (x_{i+2}-2x_{i+1}+x_{i}, S_{i}); i= 0,...n-2$\\\hline
Y Acceleration against Distance &   Acceleration relative to traveled distance in y~\cite{shen2012effectiveness}& $ (y_{i+2}-2y_{i+1}+y_{i},S_{i}), i= 0,...n$\\\hline
Avg X Acceleration against Distance & Plot of average movement acceleration with respect to culminating traveled distance by sampling~\cite{shen2012effectiveness}& $ (x_{i+2}-2x_{i+1}+x_{i})_{avg}, S_{i}; i= 0,...n-2$\\\hline
Avg Y Acceleration against Distance &  Plot of average movement acceleration with respect to culminating traveled distance by sampling~\cite{shen2012effectiveness}& $(y_{i+2}-2y_{i+1}+y_{i})_{avg}, S_{i}; i= 0,...n-2$\\\hline 
Tangential Velocity ($V_{i}$) & Velocity along  a curve~\cite{gamboa2003identity, sayed2009static, carneiro2015using} & $ 
\sqrt{ V_{x}^2 + V_{y}^2}$ \\\hline

Tangential Acceleration ($\dot V$) & How the tangential velocity of a point on a curve changes relative to time~\cite{gamboa2003identity}& $\dot V = \frac{V_{i}}{t}\hspace{0.1cm} or  \mathit{A_{i}} = \sqrt{ A_{x}^2 + A_{y}^2}$\\
\hline
Tangential Jerk ($\ddot V$) & Change of mouse acceleration on curve  relative to time~\cite{feher2012user, gamboa2003identity} &  $\frac{\dot V_{2}-\dot V_{1}}{t_{2}-t_{1}}$\\\hline
Angle of Movement ($\theta$) & Angle of the path tangent to the x axis~\cite{feher2012user} & $ \arctan\frac{y_{2}-y_{1}}{x_{2}-x_{1}}$ \\\hline
Angular Velocity ($\omega_{i}$) &  Rate of change of an angle in a circle~\cite{ernsberger2017web} & $\frac{V_{i}}{r}$\\\hline
Rate of Curvature & Rate in which the mouse curvature is changing direction~\cite{feher2012user} &  $\frac{\delta \kappa}{\delta {s}}$\\\hline

\end{tabular}
\end{table}

\begin{table}
\begin{tabular}{|p{35mm}|p{55mm}|p{45mm}|}

\hline

Total Angles, Bending Energy & Summation of  angles on  trajectory~\cite{antal2019intrusion,hinbarji2015dynamic} & $ \sum_{i=1}^{n-1}\theta_{i}, \sum_{i=1}^{n-1}|\theta_{i}|$\\\hline

Regularity        & How regular the distances between points of a curve and its geometric center~\cite{hinbarji2015dynamic}                                                    &      $\frac{\mu_{d}}{\mu_{d}+ \sigma_{d}} \in [0,1]$        \\ \hline

Trajectory of Center of Mass (TCM)& Mean time during mouse movement where  weights are given by traveled distance~\cite{feher2012user}&$ \frac{1}{S_{n}}\sum_{i=1}^{n-1} t_{i+1}D_i$\\\hline





Scattering Coefficient (SC) & Deviation of mouse movement from the movement center of mass~\cite{feher2012user} & $\frac{1}{S_{n}}\sum_{i=1}^{n-1} {t_{i+1}}^2D_i- TCM^2$\\\hline
Curvature Velocity $(V_{curve})$ & Average velocity of  curvature~\cite{feher2012user} &      $\frac{\ddot V}{(1+\dot V^2)^{3/2}}$\\\hline



Central Moments $(\mu_{n})$   & Moments around the mean  which provides  rough idea about  shape of  curve~\cite{hinbarji2015dynamic}  &   $\frac{\sum_{i=1}^{n-2}(\kappa_{i}-\mu )^{n}\left | \theta_{i} \right |}{\sum_{i=0}^{n-2}\left | \theta_{i} \right |}$            \\ \hline
Self-Intersection & Number of  points of a curve intersecting with itself~\cite{hinbarji2015dynamic}&$\sum_{i=0}^{n-1}x_{i}, y_{i}$\\\hline
Angle Feature ($\gamma$) (Law of Cosines) & Angle between the mouse's current location and two click points on squares~\cite{aksari2009active} & $ cos^{-1}\left[\frac{a^2+b^2-c^2}{2ab}\right]$\\\hline
Acceleration Beginning Time & Acceleration Time for the Beginning Segment~\cite{hassan2022intelligent}&$t_1-t_0$\\\hline
Skewness & Asymmetry of mouse sequence distributions compared to a normal distribution; also known as the third moment~\cite{ernsberger2017web, feher2012user} &$\frac{\frac{1}{n} \sum_{i=1}^{n} (X_i- \bar{X})^3}{\sqrt{\frac{1}{n-1} \sum_{i=1}^{n} (X_i-\bar{X})^2}}$ \\\hline
Kurtosis          & How much tails of mouse sequence distributions differ from the tails of a normal distribution; also known as the fourth moment &    $\frac{\frac{1}{n}\sum_{i=1}^{n}(X_{i}-X)^{4}}{\sqrt{\frac{1}{n}\sum_{i=1}^{n}((X_{i}-X)^{2})^{2}}}$            \\ \hline
\end{tabular}
\end{table}

\subsection{Public Datasets}

\begin{table}[]

\caption{Publicly Available Mouse Datasets }

\label{database_1}

\begin{tabular}{|p{4.5cm}|p{3cm}|p{1.5cm}|p{4cm}|}

\hline
{\textbf{Authors (dataset)}} & {\textbf{User Tasks}} & {\textbf{\#Subjects}} &{ \textbf{Amount of Data}}\\
\hline\hline

{Shen et al.~\cite{shen2014performance} (Chaoshen-1)} & {Fixed Static Sequence of Actions} & {58} & {17.4k samples per user}\\\hline
{Shen et al.~\cite{shen2012continuous} (Chaoshen-2)} & {App Agnostic Semi-Controlled} & {28} & {{90k mouse actions over 30 sessions} }\\\hline
{Ahmed and Traore~\cite{Dataset} (ISOT)} & {Completely Free} & {48} & {45 sessions per user for 9 weeks (284 hours of raw data)}\\\hline
{Balabit~\cite{balabit_2016}} & {Completely Free} & {10} & {Train set contains avg 937 actions and test set contains avg 50 actions}\\\hline
{Antal and Denes-Fazakas~\cite{antal2019user} (DFL)} & {Completely Free} & {21} & {1k mouse actions}\\\hline
{Belman et al.~\cite{belman2019insights} (BB-MAS)} & {Fixed Static Sequence of Actions} & {117} & {Unknown for mouse data}\\\hline
{Harilal et al.~\cite{harilal2017twos} (TWOS)} & {App Restricted Continuous} & {24} & {320 hours of active participation that included 18 hours of imposter data and at least two instances of insider attack data}\\\hline
{Siddiqui, Dave and Saliya~\cite{siddiqui2021continuous}} (Minecraft Dataset) & {Fixed Static Sequence of Actions} & {10} & {20 minutes of raw data per user}\\\hline

\end{tabular}

\end{table}

Although public datasets for mouse dynamics are not as common as keystroke dynamics, there are some publicly available mouse datasets. It is often difficult for researchers to find and compare the works of others because of the lack of standard in data collection. By accepting certain standards during data collection process will help researchers to examine certain algorithms and approaches distinctly. It will also reduce exact replication of the same effort and save time from the perspective of a researcher and the scientific community~\cite{obaidat1994multilayer}.

The datasets mentioned in the Table~\ref{database_1} of the first row are a supplement to the paper in Shen et al.~\cite{shen2014performance}. This dataset was collected based on a fixed static sequence of actions. Using this dataset, they were able to meet the European standard for commercial biometrics technology~\cite{European_standard} if taking a prolonged time for authentication. In the second row, the dataset produced by Shen et al.~\cite{shen2012continuous} was collected for continuous authentication in an app agnostic semi-controlled way. In the third row, the dataset developed by Ahmed and Traore~\cite{ahmed2007new,Dataset} under the BioTracker project is completely free, but needs prior approval from the University of Victoria  to be used for further research. In the fourth row, the Balabit~\cite{balabit_2016} dataset contains timing and positioning information of mouse pointers. It can be used to authenticate users and also test out performance of different ML algorithms. In the fifth row, Antal and Denes-Fazakas~\cite{antal2020mouse, antal2021sapimouse} described their own developed dataset called DFL, which is very similar to the Balabit dataset in terms of raw data and also completely free. In sixth row, BB-MAS~\cite{belman2019insights} is a multi-modal large dataset that includes typing, gait, mouse and swipe performed by the same user in a fixed static sequence of actions. In row 7, TWOS (the Wolf of SUTD)~\cite{harilal2017twos}, dataset was based on a multiplayer game that collects user's mouse behavioral data on simulated interactions with the system making it app restricted continuous data collection. In the last row, Siddiqui, Dave and Saliya~\cite{siddiqui2021continuous} developed a fixed static sequence of actions based dataset using Minecraft video game.

\begin{table}[]
\caption{Mouse Dynamics  based on Statistical and Pattern Recognition Algorithms}
\subcaption{\small{Gradient Boosting Machines (GBM), Support Vector Machine (SVM), Naive Bayes (NB), K-Nearest Neighbors (KNN), Decision Tree (DT), Random Forest (RF), Distance Metric (DM), Euclidean Distance (ED), Edit Distance (EDD), Error Distance (ERRD), Manhattan Distance (MD), Weibull Distribution (WBD), False Acceptance Rate (FAR), False Rejection Rate (FRR), Equal Error Rate (EER), Area under the Curve (AUC), Average Number of Genuine Actions (ANGA), Average Number of Imposter Actions (ANIA), Not Applicable (NA), Average (Avg), Accuracy (ACC), Cross Validation (CV)}, MM (Mouse Movement), DD (Drag \& Drop), PC (Point \& Click)}

\begin{tabular}{|p{2.5cm}|p{2cm}|p{.5cm}|p{2.5cm}|p{1.5cm}|p{2.5cm}|p{1.1cm}|}

\hline
\small{\textbf{Authors}} & \small{\textbf{User Tasks}} & \small{\textbf{\# Subjects}} &\small{ \textbf{Amount of Data}} & \small{\textbf{Classifier}} & \small{\textbf{Performance Metrics}}& \small{\textbf{Training and Testing}}\\
\hline\hline

\small{Revett et al.~\cite{revett2008survey}} & \small{Fixed Static Sequence of Actions} & \small{6} & \small{Click Duration (100 samples)}& \small{DM based on +/- 1.5 standard deviation} & \small{FAR (1-4\%) and FRR (1-3\%)} & \small{80/20 split}\\\hline
\small{Bours and Fullu~\cite{bours2009login}} & \small{App Restricted Continuous} & \small{28} & \small{MM (Avg of 45 sessions per user)} & \small{EDD} & \small{EER (40.1\%)} & \small{50/50 split}\\\hline
\small{Zheng, Paloski and Wang~\cite{zheng2011efficient}} & \small{Completely Free and App Agnostic Continuous} & \small{30 and 1k}  & \small{A total of 81,218 PC actions, with  average 5,801  actions per user} & \small{SVM} & \small{FRR (0.86\%) FAR (2.96\%) in 25 clicks} & \small{50/50 split}\\\hline
\small{Gamboa and Fred~\cite{gamboa2003identity}} & \small{App Restricted Continuous} & \small{25}  & \small{MM (5 hours of interaction / 180 strokes per user)} & \small{WBD} & \small{Mean EER (0.005) with std (0.001) for 100 strokes} & \small{50/50 split}\\\hline

\small{Mondal and Bours~\cite{mondal2013continuous}} & \small{Completely Free} &\small {49}& \small{MM, PC and DD (5000 samples)}  & \small{SVM} & \small{ANGA (NA), ANGA (94)}  & \small{50/50 split} \\\hline
\small{Shen, Cai and Guan~\cite{shen2012continuous}} & \small{App Agnostic Semi-Controlled} & \small{28} & \small {MM, PC, DD, C (90k mouse actions over 30 sessions)}  & \small {1-class SVM} & \small {FAR (0.37\%), FRR (1.12\%)}  & \small{1-Class Classification}\\\hline

\small {Shen et al.~\cite{shen2012effectiveness}} & \small {Fixed Static Sequence of Actions} & \small {26} & \small {MM and PC (300 samples per user)}  & \small {KNN, NN, SVM} & \small {EER (2.64\% in 110 sec})  &\small{1-Class Classification}\\\hline
\small{Shen et al.~\cite{shen2012user}} & \small{Fixed Static Sequence of Actions} & \small{37} & \small {MM and PC (5550 samples)}  & \small {1-class SVM} & \small {FAR (8.74\%) and FRR (7.96\%) in 11.8 sec}  &\small{1-Class Classification}\\\hline
\small{Shen et al.~\cite{shen2014performance}} & \small{Fixed Static Sequence of Actions} & \small{58} & \small{MM and PC (17.4k samples per user)}&\small{KNN with MD \& ED, 1-class SVM (linear \& RBF), K-means, MD} & \small{EER (5.68\%) with std (4.12)} & \small{1-Class Classification}\\\hline

\small{Shen et al.~\cite{shen2017pattern}} & \small{Completely Free} & \small{159} & \small{MM, PC, DD, C (1.5m mouse operations for all users)}&\small{1 class SVM, KNN, NN} & \small{FAR (0.09\%), FRR (1\%)} & \small{1-Class Classification}\\\hline

\small{Shen et al.~\cite{shen2010hypo}} & \small{Completely Free} & \small{20} & \small{MM, PC, DD, C, S (600 sessions for all users)}&\small{SVM, NN} & \small{FAR (1.86\%), FRR (3.46\%)} & \small{50/50 split}\\\hline
\small {Ma et al.~\cite{ma2016kind}} & \small{App Restricted Continuous} & \small{10} & \small {MM and C (500 sessions)}  & \small {SVM} & \small {Accuracy (96.3\%), FAR (1.98\%), FRR (2.10\%)} & \small{5-fold CV}\\\hline
\small {Kaixin et al.~\cite{kaixin2017user}} & \small {App Agnostic Semi-Controlled} & \small {12} & \small {MM, DD, PC, S (1000 sessions)}  & \small {SVM} & \small{ FAR (8.8\%), FRR (5.5\%) in 30sec} & \small{70/30 split}\\\hline

\small {Dominik et al.~\cite{ernsberger2017web}} & \small {App Agnostic Semi-controlled} & \small {11} & \small {MM, C, S (3283 instance (path) in total)}  & \small {LibSVM, ANN, DT, RF} & \small{ (RF) Avg Accuracy Rate (78.1\%)} & \small{10-fold CV}\\\hline
\small {Almalki, Roy and Chatterjee~\cite{almalki2019continuous}} & \small {Completely Free} & \small {10 (Balabit)} & \small {MM, DD, PC (Train set contains avg 937 actions (65 sessions) and test set contains avg 50 actions)}  & \small {DT, KNN, RF} & \small{{ACC (99.3\%), AUC (99.9\%)}} & \small{70/30 split}\\\hline
\small {Tan and Roy~\cite{tan2017insights}} & \small {Completely Free} & \small {10 (Balabit)} & \small {MM (Train set:  avg 937 actions (65 sessions) and test set:  avg 50 actions)}  & \small {Linear SVM} & \small{Avg EER (0.1829), avg AUC (0.86), avg FAR (0.21), avg FRR (0.0975)} & \small{5-fold CV}\\\hline

\end{tabular}

\label{Stat_paper}
\end{table}

\begin{table}[]

\begin{tabular}{|p{2.5cm}|p{2cm}|p{.8cm}|p{2.5cm}|p{1.5cm}|p{2.5cm}|p{1.1cm}|}
\hline

\small {Antal and Egyed-Zsigmond~\cite{antal2019intrusion}} & \small {Completely Free} & \small {10 (Balabit)} & \small {MM, DD and PC (Train set contains avg 937 actions (65 sessions) and test set contains avg 50 actions)}  & \small {RF} & \small{Avg EER (18.80\%), avg AUC (89.94\%)} & \small{10-fold CV}\\\hline
\small {Salman and Hameed~\cite{salman2018using}} & \small {Completely Free} & \small {48} & \small {MM, DD and PC (998 sessions from 48 users)}  & \small {Gaussian NB} & \small{Avg ACC (93.56\%), avg FRR (0.822), avg FAR (0.009), avg EER (0.08), avg AUC (0.98)} & \small{3-fold CV}\\\hline
\small {Hu et al.~\cite{hu2017deceive}} & \small {Completely Free} & \small {24} & \small {PC (5000 samples per user)}  & \small {RF, GBM, MLP, SVM and CNN} & \small{FRR (less than 1\%), FAR (less than 1\%) with 20 move-to-and-left click except CNN} & \small{70/30 split}\\\hline
\small {Aksari and Artuner~\cite{aksari2009active}} & \small {Fixed Static Sequence of Actions} & \small {10} & \small {MM (111 sessions)}  & \small {DM based on +/- 1.5 SD } & \small{FAR (5.9\%), FRR (5.9\%), EER (5.9\%)} & \small{90/10 split}\\\hline
\small {Gao et al.~\cite{gao2020continuous}} & \small {Completely Free} & \small {10 (Balabit)} & \small {MM (No mention of sample size)}  & \small {SVM, KNN} & \small{FAR (0.075), FRR (0.0664)} & \small{Unknown}\\\hline
\small{Zheng, Paloski and Wang~\cite{zheng2016efficient}} & \small{Completely Free} & \small{30} & \small{MM and PC (3160 mouse actions per user (150 hours of raw data)}&\small{SVM} & \small{FAR (0.86\%) and FRR (2.96\%)} after 25 clicks & \small{50/50 split}\\\hline

\small{Antal, Fejer and Buza~\cite{antal2021sapimouse}} & \small{Fixed Static Sequence of Actions} & \small{120} & \small{MM (52 blocks (MM into fixed size) per user)}&\small{CNN and 1-class SVM} & \small{AUC (0.94)} & \small{75/25 split}\\\hline

\small{Pusara and Broadley~\cite{pusara2004user}} & \small{App Restricted Continuous} & \small{18} & \small{7.6k unique cursor locations}&\small{DT} & \small{Avg FAR (0.43\%), avg FRR (1.75\%)} & \small{75/25 split}\\\hline
\small{Antal and Denes-Fazakas~\cite{antal2019user}} & \small{Completely Free  (Balabit, DFL)/App Agnostic Semi-controlled (ChaoShen)} & \small{21} & \small{MM, DD and PC (1k mouse actions)}&\small{RF} & \small{Avg AUC (0.9922) with std (0.0061)} & \small{70/30 split}\\\hline

\small{Jorgensen and Yu~\cite{jorgensen2011mouse}} & \small{App Restricted Continuous} & \small{17} & \small{MM, DD, PC for~\cite{gamboa2003identity} and MM for~\cite{ahmed2007new} (325 actions per user)}&\small{NN, Logistic Regression} & \small{Avg FAR (21\% with std 14.3\%) and avg FRR (21.5\% with std 13.4\%)~\cite{gamboa2003identity}. Avg FAR (30.3\% with std 9.8\%) and avg FRR (37.1\% with std 17.7\%)~\cite{ahmed2007new}} & \small{60/40 split}\\\hline
\end{tabular}

\end{table}

\begin{table}

\begin{tabular}{|p{2.5cm}|p{2cm}|p{.8cm}|p{2.5cm}|p{1.5cm}|p{2.5cm}|p{1.1cm}|}

\hline


\small{Carniero et al.~\cite{carneiro2015using}} & \small{Fixed Static Sequence of Actions} & \small{53} & \small{MM, DD and S (two datasets with 2.4k and 162 instances)}&\small{NB} & \small{ACC (86.4\%)} & \small{10-fold CV}\\\hline
\small{Ancien et al.~\cite{acien2020becaptcha}} & \small{Fixed Static Sequence of Actions} & \small{58} & \small{MM and PC (15K samples)}&\small{RF} & \small{Avg ACC (93\%) on one mouse trajectory} & \small{70/30 split}\\\hline
\small{Kaminsky, Enev and Andersen ~\cite{kaminsky2008identifying}} & \small{Fixed Static Sequence of Actions} & \small{15} & \small{MM (Unknown samples)}&\small{SVM, KNN} & \small{ACC (93\%)} & \small{10-fold CV}\\\hline
\small{Nakkabi, Traore and Ahmed ~\cite{nakkabi2010improving}} & \small{Completely Free} & \small{48} & \small{MM, DD and PC (2184 sessions)}&\small{Fuzzy Clustering Method} & \small{FAR (0\%), FRR (0.36\%)} & \small{1-hold out CV}\\\hline
\small{Schulz~\cite{schulz2006mouse}} & \small{Completely Free} & \small{72} & \small{MM (3.6k mouse curves)}&\small{ED} & \small{EER (11.1\%)} & \small{1 to 1 comparison}\\\hline
\small{Hashia, Pollett and Stamp~\cite{hashia2005using}} & \small{Fixed Static Sequence of Actions} & \small{15} & \small{MM (15-minute block per user)}&\small{DM based on +/- 1.5 SD } & \small{EER (15\%)} & \small{15 minutes/last ten states}\\\hline
\small{Syukri, Okamoto and Mambo~\cite{syukri1998user}}& \small{Fixed Static Sequence of Actions} & \small{21} & \small{MM (100 signatures)}&\small{ERRD points within threshold (50 pixel)} & \small{ACC (93\%)} & \small{75/25 split}\\\hline
\small {Tan et al.~\cite{tan2019adversarial}} & \small {Completely Free and App Restricted Continuous} & \small {{10 (Balabit) and 20 (TWOS)}} & \small {MM (Balabit: Train and test: avg 937 actions (65 sessions)/avg 50 actions) (TWOS: 320 hours of active participation that included 18 hours of imposter data and at least two instances of insider attack data)}  & \small {SVM, 1D-CNN, 2D-CNN}& \small{{2D-CNN (Balabit) Baseline AUC 0.96 and EER 0.10; 2D-CNN (TWOS) Baseline AUC 0.93 and EER 0.13}} & \small{80/20 split}\\\hline

\end{tabular}

\end{table}

\begin{table}

\caption{Deep Learning-based Mouse Dynamics}
\subcaption{\small{Neural Network (NN), Convolutional Neural Network (CNN), Long Short Term Memory (LSTM)}, MM (Mouse Movement), DD (Drag \& Drop), PC (Point \& Click)}
\label{CNN}

\begin{tabular}{|p{2.5cm}|p{2cm}|p{.8cm}|p{2.5cm}|p{1.5cm}|p{2.5cm}|p{1.1cm}|}

\hline
\small{\textbf{Authors}} & \small{\textbf{User Tasks}} & \small{\textbf{\# Subjects}} &\small{ \textbf{Amount of Data}} & \small{\textbf{Classifier}} & \small{\textbf{Performance Metrics}}& \small{\textbf{Training and Testing}}\\
\hline\hline

\small{Ahmed and Traore~\cite{ahmed2007new}} & \small{Completely Free} & \small{22} & \small{MM, DD and PC (45 sessions per user for 9 weeks (284 hours of raw data))}&\small{NN} & \small{FAR (2.4649\%) and FRR (2.4614\%)} & \small{1-hold out CV}\\\hline
\small{Sayed and Traore~\cite{sayed2013biometric}} & \small{Fixed Static Sequence of Actions} & \small{30}  & \small{MM (4350 samples of gesture templates over 174 sessions)}  & \small{Learning Vector Quantization (LVQ) NN} & \small{FRR (4.59\%) FAR (5.26\%)}  & \small{80/20 split}\\\hline
\small {Shen et al.~\cite{shen2009feature}} & \small {App Agnostic Semi-Controlled} & \small {10} & \small {MM, C (300 sessions in total)}  & \small {NN} & \small {FAR (0.55\%), FRR (3.0\%)}  &\small{1-hold out CV}\\\hline

\small {Hu et al.~\cite{hu2019insider}} & \small {Completely Free} & \small {10 (Balabit)} & \small {MM, DD, C and S (36k images)}  & \small {CNN (7 Layers)} & \small{FAR (2.94\%} and FRR (2.28\%) & \small{85/15 split}\\\hline

\small {Chong et al.~\cite{chong2019user}} & \small {Completely Free (Balabit) and App Restricted Continuous (TWOS)} & \small {{10 (Balabit) and 20 (TWOS)}} & \small {MM ((Balabit) Train and test sets (avg 937 actions (65 sessions)/avg 50 actions) (TWOS) 320 hours of active participation that included 18 hours of imposter data and at least two instances of insider attack data)}  & \small {SVM, LSTM, CNN-LSTM 1D-CNN, 2D-CNN}& \small{{2D-CNN (Balabit) Baseline Avg AUC 0.96 and Avg EER 0.10; 2D-CNN (TWOS) Baseline AUC 0.93 and EER 0.13}} & \small{5-fold CV}\\\hline

\small{Antal and Fejer~\cite{antal2020mouse}} & \small{Completely Free} & \small{{10 (Balabit) and 21 (DFL)}} & \small{MM (370 blocks per user)}&\small{1D CNN} & \small{AUC (0.98)} & \small{80/20 split}\\\hline
\small{Everitt and McOwan~\cite{everitt2003java}} & \small{Completely Free} & \small{41} & \small{MM (unknown samples)}&\small{NN} & \small{FAR (4.4\%), ACC (~99\%)} & \small{4/96 split}\\\hline

\small{Hinbarji, Albatal and Gurrin~\cite{hinbarji2015dynamic}} & \small{Completely Free} & \small{10} & \small{MM, DD and PC (16.5k actions)}&\small{NN} & \small{EER (5.3\%), Authentication time (18.7 minute)} & \small{50/50 split}\\\hline
\small {Li et al.~\cite{li2016}} & \small {App Restricted Continuous} & \small {{6}} & \small {Keystroke, MM (25 minutes per subject)}  & \small {SVM} & \small{{CCR Leave One Subject Out 75.5\%; CCR All Subjects 78.9\%;}} & \small{10-fold CV}\\\hline
\small {Hamdy et al.~\cite{hamdy2011}} & \small {App Restricted Static} & \small {{274}} & \small {MM (2740 sessions)}  & \small {Weighted-Sum} & \small{{EER 2.11\%}} & \small{70/30 split}\\\hline
\small {Fu et al.~\cite{fu2022}} & \small {App Restricted Continuous} & \small {{18}} & \small {MM (11 trials per subject)}  & \small {CNN-RNN, PADTW} & \small{{CNN RNN EER 2.69\%; PADTW EER 8.53\%;}} & \small{5-fold CV}\\\hline
\small {Siddiqui et al.~\cite{siddiqui2022}} & \small {App Restricted Continuous} & \small {{40}} & \small {MM}  & \small {1D-CNN, LSTM-RNN, ANN, KNN, SVM, RF} & \small{{1D-CNN Mean ACC 0.8573; Mean FPR 0.1546; Mean F1 0.9099;}} & \small{50/50 split}\\\hline

\hline

\end{tabular}

\end{table}

\begin{table}
\begin{tabular}{|p{2.5cm}|p{2cm}|p{.8cm}|p{1.5cm}|p{2.5cm}|p{2.5cm}|p{1.1cm}|}

\hline
\small{\textbf{Authors}} & \small{\textbf{User Tasks}} & \small{\textbf{\# of Subjects}} &\small{ \textbf{Amount of Data}} & \small{\textbf{Classifier}} & \small{\textbf{Performance Metrics}}& \small{\textbf{Training and Testing}}\\
\hline\hline

\small {Kang et al.~\cite{kang2023user}} & \small {App Restricted Continuous} & \small {{60}} & \small {MM, Keystroke (5,000–10,000 KLM operators per experiment)}  & \small {Decision Tree} & \small{{AUC Photoshop 0.72-0.92; AUC 3DS Max 0.84-0.87;}} & \small{NA}\\\hline
\small {Lopez et al.~\cite{lopez2023adversarial}} & \small {App Restricted Continuous} & \small {{58}} & \small {Balabit, TWOs, Typing-BD}  & \small {SVM, RF, DNN} & \small{{Mean AUC SVM 0.84; Mean AUC RF 0.86; Mean AUC DNN 0.86;}} & \small{70/30 split}\\\hline
\small {Hassan et al.~\cite{hassan2022intelligent}} & \small {Fixed Static Sequence of Actions} & \small {{50}} & \small {MM (3859 total actions)}  & \small {LightGBM, XGBoost, Logistic Regression, Random Forest, SVC, KNN} & \small{{ACC LightGBM 0.92; ACC XGBoost 0.88; ACC Logistic Regression 0.77; ACC Random Forest 0.87; ACC SVC 0.73; KNN 0.68;}} & \small{70/30 split}\\\hline
\small {Shi et al.~\cite{shi2023user}} & \small {Completely Free} & \small {{41}} & \small {MM, Keystroke (4 hours per user)}  & \small {AAN, NN, LR, SVM, DT KNN, NB} & \small{{ACC AAN 89.22; ACC SVM 71.6496; ACC LR 74.5226; ACC DT 68.4263; ACC NN 76.5779; ACC NB 56.4139; ACC KNN 65.0020}} & \small{5-fold CV}\\\hline
\small {Abin et al. \cite{abin2023continuous}} & \small {Completely Free} & \small {{12}} & \small {MM, Keystroke}  & \small {OCSVM-PSO, SVM, HNS, RF, NB, K-Means Clustering} & \small{{ACC SVM 34.68\%; ACC HNS 88.29\%; ACC RF 96.48\%; ACC NB 88.29\%; ACC K-Means Clustering 52.01\%; ACC OCSVM-PSO 97.83\%;}} & \small{80/20 split}\\\hline
\end{tabular}
\end{table}

\section{Machine Learning Algorithms for Mouse Dynamics}
\label{State_Based_Papers}
In this section, we first classify  mouse dynamics research  
based on statistical and pattern recognition algorithms (Table~\ref{Stat_paper}). 
Statistical algorithms can be defined as mean, standard deviation, minimum, or maximum, to more complex methods such as T-Test, Euclidean Distance, Manhattan distance, P-Value Test and so on~\cite{banerjee2012biometric}. Conversely, pattern recognition algorithms use various features to find patterns, in order to classify them into different groups, for example, SVM, KNN, GBM, etc.~\cite{theodoridis2010introduction}. 
Based on their contributions, we further group these statistical and pattern recognition algorithms  into three categories: 
1) Feature Selection (papers that are classified solely based on their feature selection methods as a primary reason to authenticate) 2) Performance Evaluation (papers fall in this category on the sole basis for performance evaluation of different classifiers or models) 3) Spoof Attack (papers that explain mouse authentication to be evaluated against spoof attacks).

\subsection{\textbf{Feature Selection}}
In mouse dynamics research, several feature selection techniques have been explored to improve authentication systems. Zheng, Paloski, and Wang~\cite{zheng2011efficient, zheng2016efficient} focus on angular-based metrics such as direction, angle of curvature, and curvature distance, making them environment-independent features. Gamboa and Fred~\cite{gamboa2003identity} authenticate users based on mouse movement strokes using sequential forward selection (SFS) for feature isolation. Shen et al.\cite{shen2010hypo} proposed two feature selection techniques using sequential forward selection (SFS) and Plus-M-Minus-R (PMMR). They found that PMMR with SVM provided superior results in authentication. They later used the \textit{PrefixSpan} algorithm for feature selection from frequent mouse activities, achieving stable mouse patterns on a fixed static sequence of actions dataset\cite{shen2012continuous}. In a later study by Shen et al. ~\cite{shen2017pattern}, a similar pattern-growth based mining method was used to extract features from consistent behavioral segments with a completely free dataset, demonstrating the stability of feature selection methods. Hamdy et al. \cite{hamdy2011} incorporated principles of visual search capability and short-term memory effect into a static biometric authentication system using mouse dynamics. They demonstrated that visual search capability and short term memory were very important features that significantly bolstered performance of their mouse dynamics system. These methods all examined different features that can be used to bolster performance in different datasets. Some features may have better performance with a specific type of dataset, but some in the case of Shen et al. \cite{shen2012continuous, shen2017pattern} are shown to work on both fixed static sequence of actions and completely free datasets.

\subsection{\textbf{Performance Evaluation}}
In mouse dynamics, there have been several studies pertaining to the performance evaluation of different classifiers. These studies utilized many different types of datasets, and demonstrated the performance of many different classifiers in regards to them. It is important to note that the best performing classifier varied by data collection type as well as the features extracted from the raw data. Unfortunately, benchmarking studies based on common datasets do not exist.

Jorgensen and Yu~\cite{jorgensen2011mouse} investigated two work (Ahmed and Traore~\cite{ahmed2007new} and Gamboa and Fred~\cite{gamboa2003identity}) by mimicking their experiments to determine limitations of their mouse authentication system and evaluate performances. At first, they tried to determine if the environment was tightly controlled how effective a mouse was. Secondly, if the enrollment and verification data were collected in different computing environments how effective it was to authenticate a user. Gamboa and Fred's method performed better in the first experiment, and in the second experiment, average error rates rose for both cases (Ahmed and Traore, and Gamboa and Fred) when training and testing data were collected on two different devices. They recommended that one way to improve error rates is to reduce the noise in the raw data and apply fusion techniques. Conversely, Mondal and Bour~\cite{mondal2013continuous} used six classifiers for their experiments to evaluate performances. Among them, Support Vector Machine (SVM) performed the best. In a similar manner but with 17 classifiers, Shen et al.~\cite{shen2014performance} established a public dataset for the sake of enhancing research in this area. The Nearest Neighbor with Manhattan using 200 samples gives the best result for authentication. They also explore scalability of the system by ranging users from 2 to 58. Furthermore, they found that at around 22 users, EER becomes stable as a result of adequate genuine users in the authentication system. Almalki, Roy and Chatterjee~\cite{almalki2019continuous} investigated a practical evaluation of different ML algorithms, also conducted on an open source dataset (ISOT)~\cite{ahmed2010mouse}. Among all the classifiers tested, they found random forest (RF) provided the best AUC and ACC. Tan and Roy~\cite{tan2017insights} investigated the performance of different curve smoothing techniques on the mouse movement sequences for an authentication model. They used three time-series forecasting models: cubic spline, AR (Auto Regressive) Model and lastly, AR with Moving Average (ARMA). They found AR curve fitting technique with the RF classifier showed the superior result. Siddiqui et al. \cite{siddiqui2022} expanded on their earlier public Minecraft dataset \cite{siddiqui2021continuous} by increasing the number of users from 10 to 40 in an attempt to provide the research community with more naturalistic mouse dynamics data. They also sought to give baselines of performance of several classifiers (1D-CNN, LSTM-RNN, ANN, KNN, SVM, RF) for the wider research community and attempted to discover a best performing classifier for mouse dynamics. After analysis of each classifier, it was found that the deep learning models outperformed the machine learning models in accuracy, with ANN performing the best out of all of them. Hassan et al. \cite{hassan2022intelligent} presented a mouse dynamics model that they used for detecting different mouse users for anti-cheat. 39 different features were extracted from raw mouse movement data across 2 trials. The performance of several different classifiers was tested for this purpose (KNN, SVC, RF, Logistic Regression, XGBoost, and LightGBM). LightGBM was shown to have the best performance, followed by XGBoost, RF, Logistic Regression, SVC, and KNN. Abin et al. \cite{abin2023continuous} presented an anomaly detection method for authenticating users using OCSVM (One Class SVM) with PSO (Particle Swarm Optimization) on a completely free non application constrained dataset. This proposed method was compared against multiple other classifiers, including one unsupervised model (OCSVM-PSO, SVM, HNS, RF, NB, K-Means Clustering). The best performing method was OCSVM-PSO, followed by RF, NB and HNS, K-Means Clustering, and SVM. 

\subsection{\textbf{Spoof Attack}}
In mouse dynamics, many studies have been done to determine how the modality could be spoofed, and what methods performed best for doing so, as well as defenses against these methods. Dominik et al.~\cite{ernsberger2017web} proposed a mouse behavioral dynamics visualization tool that could be used for forensic purposes to gather and store digital information for any cyber security violations. Their tool worked on several news agency websites to collect mouse data from participants and provided a defense mechanism against spoof attacks. Also in a forensic manner, Hu et al.~\cite{hu2017deceive} proposed a mouse movement simulation method to inspect the vulnerability of the existing authentication methods. They created synthetic and simulation data on par with real data, which were moving tracks consisting of points along with timestamps, after identifying different objects (e.g., icons of certain applications) as coordinate representations. Finally, they showed that synthetic data works almost the same as real data and better than simulation data to inspect the vulnerability of existing authentication method. Conversely, Antal and Egyed-zsigmond~\cite{antal2019intrusion} explored mouse dynamics in light of intrusion detection (i.e. a form of spoof attack). They measured how many mouse actions (MM, PC, DD)  are needed to get an optimum result to detect an intrusion and which types of mouse actions are the most important ones. From that dataset, they determine that MM and PC performed almost equally and AUC became 1 after a set of actions. Tan et al.~\cite{tan2019adversarial} developed a threat model based on prior knowledge of how attackers can bypass an authentication system. They explore three different attack strategies: 1) statistics-based (i.e., assuming that the attacker has access to the recorded target user's data) 2) imitation-based (i.e., trained model produce mouse trajectories to mimic a user's mouse movement sequence also known as teacher-forcing approach, which is based on recurrent neural network~\cite{williams1989learning}), and 3) surrogate-based (i.e., train a substitute classifier assuming that it learns the same functionalities as the target classifier from a substitute dataset and performs a white-box attack starting with random mouse movement sequences and alter them with some constant repetitions~\cite{papernot2017practical}). The authors showed that imitation or surrogate based attacks performed better than a statistical approach. Lopez et al. \cite{lopez2023adversarial} focused on an attack technique for behavioral biometric systems based on reusing genuine user inputs and reapplying them in order to impersonate a user on a mouse and keystroke dynamics protected system. Two approaches were used to perform attacks, 1) SCRAP (Synthetically Composed Replay Attack Procedure) and 2) Adversarial Black-Box Attacks. They tested against 3 different machine learning classifiers, SVM, RF, and DNN. The performance of SCRAP against these three classifiers was higher than that of adversarial machine learning. In order to limit the effectiveness of SCRAP attacks, they also proposed employing adversarial training by training with attack samples, which they found to be a well performing countermeasure. Instead of focusing on just features to improve classification, Fu et al. \cite{fu2022} developed a continuous mouse dynamics authentication system that operated around the principle of ``induced expertise''. Induced expertise is the notion that if a user interacts with a modified version of a system in their daily tasks, they will collect a certain level of expertise on that system over an untrained user. Two classifiers were used, CNN-RNN and PADTW (an improved version of the Dynamic Time Warping algorithm), with CNN-RNN yielding better performance in every scenario. They found that an inexperienced attacker is more distinguishable than an experienced attacker, but even in the case of an experienced attacker, performance is still excellent with their system.

\subsection{\textbf{Fusion of Mouse Dynamics with Other Modalities}}

Li et al. \cite{li2016} developed a user independent model to recognize human attention level. They used three different modalities in order to classify a participants attention level in their data: 1) Facial Expression 2) Eye Gaze 3) Mouse Dynamics. They extracted 7 total mouse features from their raw data, 9 eye gaze features, and 80 facial expression features. They used the wrapper method with a best-first searching approach based on a linear SVM for classification in order to reduce the number of features and to use only important features. They found that mouse dynamics was the worst performing modality for their data, but still offered improvement over their baseline, and that facial expression was their best performing modality. They concluded that human attention level could successfully be classified best by the fusion of the three modalities studied. The poor performance of mouse dynamics compared to other modalities is likely linked to the low feature count, and what features were extracted. Mouse dynamics has been shown to perform at a very high level in other studies we surveyed. However, it is worth noting that the best performance achieved in this paper was through the fusion of the three modalities.

\subsection{Deep Learning Based Mouse Dynamics}
\label{CNN_Based_Papers}
In this section, we review  the literature  on mouse dynamics  based on deep learning algorithms (i.e., artificial neural network (ANN), which is based on the concept of how biological neurons function together in the human brain to be implemented as a non-linear data modeling tool)~\cite{obaidat1993online}. An early example of this research is a study by Everitt and McOwan~\cite{everitt2003java}, who proposed a Java based online behavioral biometric authentication system using keystrokes and mouse signatures. The system mainly authenticated using NN in three phases: registration, training and testing. The participants were asked to register and train with their own information, and also provided a test set of forged samples for a random selection of other users via a Web-based applet remotely. A later study by Ahmed and Traore~\cite{ahmed2007new} developed a different novel technique that modeled human behavior from captured data and classified it using a three-layer feed forward perceptron Neural Network (NN). They created a concatenation of 39 features from the set of factors and used them as inputs to the NN. The NN gives the same importance to all the features simultaneously as a deciding factor for authentication. For the training phase, they used an NN to train each user and keep the trained network stored in a database. In the testing phase, they loaded the genuine user's stored NN network and compared it with the confidence ratio (CR) with all the users to determine the similarity between two behaviors. Shen et al.~\cite{shen2009feature} addressed the issue of mouse variability by pre-processsing features using principal component analysis (PCA) and feature space analysis, using manifold learning called ISOMAP (isometric mapping)) to reduce dimensionality. They then utilized NN for classification to get the best authentication results~\cite{jaadi,gupta_2020}). Sayed et al.~\cite{sayed2013biometric} demonstrated a new mouse dynamics framework using mouse gestures for static authentication. Unlike Ahmed and Traore and Shen et al., Sayed et al. used a data smoothing technique (weighted least-squares regression (WLSR) method and Pierce's criterion) on their data. This was then used with an NN in order to authenticate users. Hinbarji, Albatal and Gurrin~\cite{hinbarji2015dynamic} proposed a method of authenticating users based on mouse movements using NN. To improve performance, they combined multiple curves (sessions), because a single curve did not contain enough information. Furthermore, they tested their system by increasing the size of signatures by 100, 200 and 300 curves. Chong et al.~\cite{chong2018mouse} investigated two open source datasets (Balabit and TWOS) using 2D-CNN in comparison with 1D-CNN and hand crafted features. They also presented a multi-label joint training classifier, where it predicts a set of target labels, in which each label represents a different classification problem. They used transfer learning using GoogleNet architecture, modified to a multi-label architecture. They also used their own mapping methods to convert time series data into 2D image data, where they found fused curve (i.e., mouse curves are meshed together to get a longer curve in 2D image) to provide the best result for their experiments. Hu et al.~\cite{hu2019insider} took a different approach by mapping all basic mouse actions to images and then classify them using CNN. Similarly to Chong et al., they also used the Balabit dataset, but developed their own mapping technique to map time-series data to 2D image data for the CNN input. Their results showed that they can complete user authentication in 1.78s. Similarly to Chong et al. and Hu et al., Antal and Fejer~\cite{antal2020mouse} used the Balabit and DFL datasets, but used NN model in order to learn features automatically from the raw data without any feature extraction.

\section{Widget Interactions as a Behavioral Biometric Authentication System}
\label{widget}
Widget interactions refer to what (e.g., clicking on a button, hover over an icon, moving from a window to window) and how a user interacts (e.g., duration) with the composites of a GUI system~\cite{khan2021authenticating}. Other authors also call widget interactions a GUI based authentication mechanism~\cite{pusara2007examination, bailey2014user, imsand2008applications}.  We believe that widget interactions, a new kind of modality, is closely related to mouse dynamics, due to the user's interaction with widgets (e.g., buttons, icons) via a mouse on a GUI. Retrospectively, it is also different from mouse dynamics, because widget interactions only include widgets at the time of interaction for authentication, which is not considered in mouse dynamics. Generic mouse dynamics only analyzes the mouse behavior, such as mouse movement and mouse wheeling, without providing any relevance to the widgets. The main goal of mouse dynamics is to authenticate users based on mouse movement/trajectory patterns. As a result, it can be hypothesized that widget interactions will provide many other discriminating factors that mouse dynamics does not consider.

\begin{figure*}
\centering

   \includegraphics[width=4.5in]{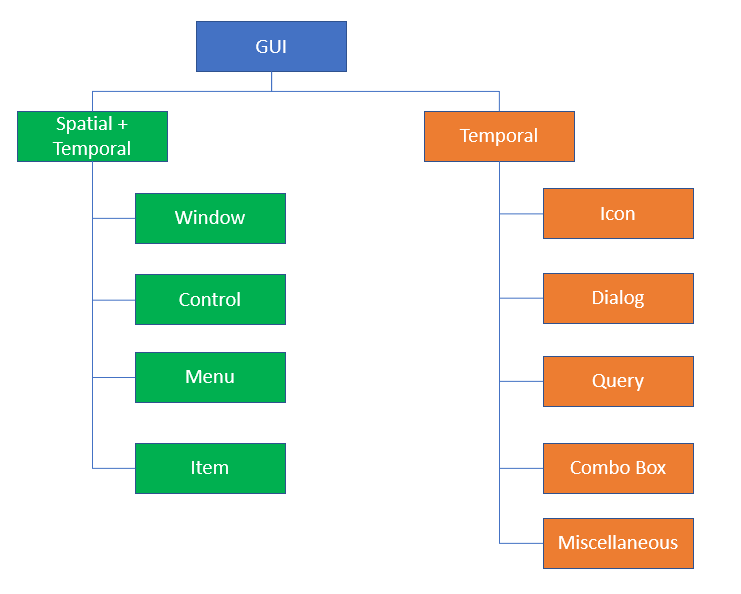}

   \caption{Pusara's Capture of GUI Events~\cite{pusara2007examination}}
   \label{pusara}

\end{figure*}

 Earlier papers that discuss a GUI based or GUI related authentication system include Pusara~\cite{pusara2007examination}, Imsand~\cite{imsand2008applications} and Bailey, Okolica and Peterson~\cite{bailey2014user}. Pusara took the next step of proposing a multi-modal re-authentication (continuous authentication) system in a close setting environment (i.e., a strictly controlled environment) with 61 users to apply supervised learning algorithms. She then examined the supervised classifier's ability to detect the unseen users as a part of her experiments. She characterized the GUI events into two categories: spatio-temporal and temporal events. The Spatio-temporal category involves both spatial and temporal activities (e.g., window (i.e., scroll bar, min/max, restore/move), control (i.e., application and process control, open/close), menu (e.g., open, select, navigate, close) and item (e.g., list, button etc.)). She calculated mean, standard deviation and skewness of distance, speed, angle, X, Y coordinates and \emph{n}-graph related to GUI events. Respectively, the temporal category involves only time features (e.g., icons, dialog, query, combo box and miscellaneous activities). She also calculates mean, standard deviation and skewness of \emph{n}-graph duration between temporal events (see Figure~\ref{pusara}). Furthermore, she used a decision tree as the final classifier for authentication. The exploratory results are encouraging with FAR (23.37\%) and FRR (1.5\%) in a  detection time of 50s, and with unseen data, the FRR is 2.25\% with a detection time of 49.3s. Imsand proposed only a GUI based authentication system on 31 users. His main goal was to authenticate users as well as protect normal users from any masquerade attacks (e.g., disgruntled employee of a company). He collected raw data using the Windows hook procedure~\cite{hickeys_2012} based on events (i.e., occurrence of an action such as left button click), object class such as edit, message recipients (e.g., message transmitted from operating system to a running process (e.g., Internet Explorer (IE)), handle (e.g., event specific information such as an ID to keep track of control). He utilized similarity scores based on count to calculate the differences between reference and unknown templates using TF-IDF (term frequency-inverse document frequency) and Jaccard Similarity index. He also experimented with static and variable cutoffs to determine an attack. Variable cutoff was used to label a session being attacked to be customized for each user. The author found that Jaccard coefficient is the most effective way to authenticate users and detect attacks with varying cutoffs of FAR (0\%) and FRR (8.66\%). Bailey, Okolica and Peterson proposed a multi-modal authentication system (keystrokes, mouse and GUI) with 31 participants. Their raw data collection for GUI system was identical to Imsand~\cite{imsand2008applications}. Next, they used two different fusion techniques: feature and decision fusions. Decision fusion provided the best FAR (2.24\%) and FRR (2.10\%) as an ensemble of all decisions.

Pusara~\cite{pusara2007examination}, Imsand~\cite{imsand2008applications} and Bailey, Okolica and Peterson~\cite{bailey2014user} all proposed a GUI based authentication system in a different way. They all collected data from users performing different tasks (navigation on different websites as a fixed static sequence of actions). Additionally, Pusara, and Bailey, Okolica and Peterson developed a multi-modal authentication system, whereas Imsand developed a system to authenticate users based on a GUI only. Moreover, their authentication methods of using different feature extraction techniques and algorithms were completely different from one another. Pusara experimented with raw data to calculate mean, standard deviation and skewness of distance, angle, speed and duration from the GUI events. On the other hand, Imsand used only a counting based method to calculate the frequency of an event (i.e., frequency of certain words to extract features using TF-IDF to determine the importance of an event to the overall corpus) and Jaccard index (i.e., similarity computation between two sets of data). Bailey, Okolica and Peterson performed two different fusion techniques to improve FAR and FRR for a multi-modal system. They used a window sliding technique in feature level fusion to utilize a chunk of feature samples from keystrokes, mouse activities and GUI at a time for authentication. Hence, these research papers provide a new outlook within the behavioral biometrics community for further explorations.

Our prior publication~\cite{khan2021authenticating} is closely related to Pusara, Imsand, and Bailey, Okolica and Peterson in terms of data collection process, primarily with respect to the GUI based authentication system. However, we only focused on Facebook to collect raw data (e.g., types of icons, hours-of-day, days-of-week) instead of different websites and word processing activities to authenticate users focusing at a granular level. At first, we invited 8 voluntary users to provide data remotely using our own developed software (a Firefox extension). After collecting raw data, we extracted features using one-hot encoding method. One-hot encoding method is a count based method similar to TF-IDF, where categorical features are assigned with Boolean value of 0 or 1. We also used trigonometric approaches to model our features continuously, such as hour-of-day and day-of-week, as an alternative representation due to the nature of them being ordinal cyclic variables. For authentication, we used two classifiers, SVM and GBM, to measure EER and ACC. Our GBM provided the best results with EER (2.75\%) and ACC (97.85\%). We also performed an ablation study to investigate the importance of features and measure how overlapping data impact the metrics. 

A more recent study by Kang et al. \cite{kang2023user} created a model based on the Keystroke Level Model (KLM) for authenticating users based on UI interaction sequences. Mouse movement and keystroke data are both captured in this study, and KLM rules were used to encode the data, which are based off of HCI principles. This essentially directly fuses the two modalities so they can be considered as one for authentication. The use of KLM also makes this authentication system non application constrained. Autodesk 3DS Max and Adobe Photoshop were used to validate the model. Overall they found model performance to be reasonably good, with performance being higher in 3DS Max than in Photoshop. This gives optimism for adopting other HCI models in mouse dynamics authentication research. This is an interesting approach because it deals with fusion of keystroke and mouse for widget dynamics interactions.

\section{Challenges and Research Opportunities in Mouse Dynamics and Widget Interactions}
\label{challenges}

Although promising results have been attained through mouse dynamics and GUI based authentication systems, in the following we discuss several remaining challenges  before such systems can be deployed. 
    
    Data quality can be severely impacted by data interoperability issues when the users use different computer/laptop/mouse during data collection. 
    According to Jain, Ross and Pankanti~\cite{ross2006handbook},  interoperability issues may arise from  different computers, or computer related equipment, having different screen resolutions, speed, memory, different brands of mice, etc. 
    Phillips et al.~\cite{phillips2000introduction} demonstrated that the effectiveness of facial recognition algorithms on different images is drastically impacted by different camera brands, but there has no investigation of this issue in mouse dynamics.
   
    As mentioned in Section~\ref{A}, we hypothesize that Fitts' and Hick's laws can be used directly towards mouse authentication and widget interactions~\cite{khan2021authenticating}. In our own prior study, a user hovered over different widgets on a Facebook platform~\cite{khan2021authenticating}. Different widgets have different text based lengths, which can be construed as widths ($W$) and trajectory distances can be considered as amplitudes ($A$). According to Fitts' law, this $W$ and $A$ along with $I_d$ and $I_p$ can be used as features for mouse authentication. Other models derived from Fitts' law could also have potential as features, either for predicted movement time versus actual, or predicted error time versus actual. Considering the various applications and dimensions of these models, there is much work needed to be done in performing performance evaluations for each model for mouse dynamics authentication. Similarly, we  can also apply Hick's law to use as a feature based on our study. For example, when a user hovers over a like button, multiple choices appear for selection. As a result, we can measure the response time it takes for a user to select a widget as a feature.

   In general psychology as a science  has been largely ignored in behavioral biometric research. We believe that human psychology can play an important role in both explaining and improving how users are authenticated. For instance, Fitts' and Hick's laws from experimental psychology have been applied in HCI research to make it more efficient, as have multiple models discussing the speed-accuracy tradeoff. However, these laws and models have not been explored by the research community for authentication. Moreover, there are many other unknown factors due to a lack of understanding based on social and behavioral sciences, such as how human emotions and societal situations can affect the system output during authentication~\cite{shneiderman1980software}.
   
    Benchmarking on public datasets is important to increase research generalizability. The mouse dynamics scientific community lacks benchmarking with public datasets to further improve their authentication techniques. Researchers need to objectively measure how well they are doing on a particular problem based on established benchmarks of a particular dataset. 
    Like other behavioral biometric modalities, more research is needed to improve our understanding of both the intra-subject and inter-subject variabilities of mouse dynamics.
    Mouse dynamics and widget interactions based on a GUI system  can be susceptible to other vulnerabilities, such as real data being replaced with synthetic data or different feature sets, hijacking the classifier to falsely allow an imposter to circumvent the system. Furthermore, data can be intercepted and replayed by an attacker with their own data (replay attack)~\cite{ratha2001analysis}.
    

While the focus of this survey was on mouse dynamics, it would be desirable to include
additional consideration to include trackpads for laptop computers. Modern laptop computers offer the speed and power of desktop computers, with the advantage of size, mobility, and convenience. Laptop computers may use trackpads opposed to the traditional mouse. It would be interesting to apply this research from data collected from a trackpad and compare the results to the physical movements of a mouse. Consideration would need to account for the continued point of origin. Meaning, when using a mouse, the device typically is not repositioned after small movements, whereas when using a trackpad, the finger moves the cursor, and is repositioned for comfort and limited range, before continuing to its destination. The length of this range movement of the finger opposed to the mouse is an interesting topic, and one that would require future investigation. Another focus would include touchscreen and handheld devices such as tablets, and cell phones. The evolution of computing devices has introduced devices beyond traditional computers with a mouse. The popularity of small devices that contain sensitive information is increasing with the adoption of cell phones and tablets. In some cases, these have become the primary devices. 

Lastly, there remains the question of how the mouse dynamics can be adopted in  real-life scenarios. Given the  authentication performance reported in the literature, it is unlikely for mouse dynamics to act as a primary authentication modality alone. Therefore, future work is needed to investigate how  mouse dynamics may comlement other identification and authentication methods. These would require formal experiments such as Wahab, Hou, and Schuckers~\cite{codaspy-usability}.
\section{Conclusion}
\label{conclusion}

In this paper, we provided a survey of state-of-the-art behavioral biometric systems that utilize mouse dynamics and widget interactions. 
Although these modalities have shown promising performance in user authentication, they are also relatively new. Therefore, there exist plenty of research opportunities for further  exploration. We began by reviewing the literature that investigates the impact of human psychology on mouse dynamics and widget interactions. We then surveyed the literature on mouse dynamics and widget interactions, along the dimensions of tasks and experimental settings for data collection, public datasets, features, algorithms (statistical, machine learning, and deep learning), and fusion and performance. Lastly, we discussed the challenges that exist when dealing with these modalities, which provided directions for future explorations.


\begin{acks}
Simon Khan and Michael Manno are also affiliated with and funded by the US Air Force Research Lab (AFRL), Rome, NY. Daqing Hou and Charles Devlen were partially supported by United States NSF award TI-2122746.
\end{acks}

\nocite{*} 




\begin{thebibliography}{146}


\ifx \showCODEN    \undefined \def \showCODEN     #1{\unskip}     \fi
\ifx \showDOI      \undefined \def \showDOI       #1{#1}\fi
\ifx \showISBNx    \undefined \def \showISBNx     #1{\unskip}     \fi
\ifx \showISBNxiii \undefined \def \showISBNxiii  #1{\unskip}     \fi
\ifx \showISSN     \undefined \def \showISSN      #1{\unskip}     \fi
\ifx \showLCCN     \undefined \def \showLCCN      #1{\unskip}     \fi
\ifx \shownote     \undefined \def \shownote      #1{#1}          \fi
\ifx \showarticletitle \undefined \def \showarticletitle #1{#1}   \fi
\ifx \showURL      \undefined \def \showURL       {\relax}        \fi
\providecommand\bibfield[2]{#2}
\providecommand\bibinfo[2]{#2}
\providecommand\natexlab[1]{#1}
\providecommand\showeprint[2][]{arXiv:#2}

\bibitem[Abin et~al\mbox{.}(2023)]%
        {abin2023continuous}
\bibfield{author}{\bibinfo{person}{Ahmad~Ali Abin}, \bibinfo{person}{Parisima Hosseini}, {and} \bibinfo{person}{Alireza Torabian~Raj}.} \bibinfo{year}{2023}\natexlab{}.
\newblock \showarticletitle{Continuous User Authentication Using a Combination of Operation and Application-related Features}.
\newblock \bibinfo{journal}{\emph{Journal of Innovations in Computer Science and Engineering (JICSE)}} (\bibinfo{year}{2023}), \bibinfo{pages}{11--27}.
\newblock


\bibitem[Abramson and Aha(2013)]%
        {abramson2013user}
\bibfield{author}{\bibinfo{person}{Myriam Abramson} {and} \bibinfo{person}{David Aha}.} \bibinfo{year}{2013}\natexlab{}.
\newblock \showarticletitle{User authentication from web browsing behavior}. In \bibinfo{booktitle}{\emph{The Twenty-Sixth International FLAIRS Conference}}.
\newblock


\bibitem[Acien et~al\mbox{.}(2020)]%
        {acien2020becaptcha}
\bibfield{author}{\bibinfo{person}{Alejandro Acien}, \bibinfo{person}{Aythami Morales}, \bibinfo{person}{Julian Fierrez}, {and} \bibinfo{person}{Ruben Vera-Rodriguez}.} \bibinfo{year}{2020}\natexlab{}.
\newblock \showarticletitle{BeCAPTCHA-Mouse: Synthetic mouse trajectories and improved bot detection}.
\newblock \bibinfo{journal}{\emph{arXiv preprint arXiv:2005.00890}} (\bibinfo{year}{2020}).
\newblock


\bibitem[Adeoye(2010)]%
        {adeoye2010survey}
\bibfield{author}{\bibinfo{person}{Olufemi~Sunday Adeoye}.} \bibinfo{year}{2010}\natexlab{}.
\newblock \showarticletitle{A survey of emerging biometric technologies}.
\newblock \bibinfo{journal}{\emph{International Journal of Computer Applications}} \bibinfo{volume}{9}, \bibinfo{number}{10} (\bibinfo{year}{2010}), \bibinfo{pages}{1--5}.
\newblock


\bibitem[Ahmed and Traore(2007)]%
        {ahmed2007new}
\bibfield{author}{\bibinfo{person}{Ahmed Awad~E Ahmed} {and} \bibinfo{person}{Issa Traore}.} \bibinfo{year}{2007}\natexlab{}.
\newblock \showarticletitle{A new biometric technology based on mouse dynamics}.
\newblock \bibinfo{journal}{\emph{IEEE Transactions on dependable and secure computing}} \bibinfo{volume}{4}, \bibinfo{number}{3} (\bibinfo{year}{2007}), \bibinfo{pages}{165--179}.
\newblock


\bibitem[Ahmed and Traore(2010)]%
        {ahmed2010mouse}
\bibfield{author}{\bibinfo{person}{Ahmed Awad~E Ahmed} {and} \bibinfo{person}{Issa Traore}.} \bibinfo{year}{2010}\natexlab{}.
\newblock \showarticletitle{Mouse dynamics biometric technology}.
\newblock In \bibinfo{booktitle}{\emph{Behavioral Biometrics for Human Identification: Intelligent Applications}}. \bibinfo{publisher}{IGI Global}, \bibinfo{pages}{207--223}.
\newblock


\bibitem[Aksari and Artuner(2009)]%
        {aksari2009active}
\bibfield{author}{\bibinfo{person}{Yigitcan Aksari} {and} \bibinfo{person}{Harun Artuner}.} \bibinfo{year}{2009}\natexlab{}.
\newblock \showarticletitle{Active authentication by mouse movements}. In \bibinfo{booktitle}{\emph{2009 24th International Symposium on Computer and Information Sciences}}. IEEE, \bibinfo{pages}{571--574}.
\newblock


\bibitem[Almalki et~al\mbox{.}(2019)]%
        {almalki2019continuous}
\bibfield{author}{\bibinfo{person}{Sultan Almalki}, \bibinfo{person}{Prosenjit Chatterjee}, {and} \bibinfo{person}{Kaushik Roy}.} \bibinfo{year}{2019}\natexlab{}.
\newblock \showarticletitle{Continuous authentication using mouse clickstream data analysis}. In \bibinfo{booktitle}{\emph{International Conference on Security, Privacy and Anonymity in Computation, Communication and Storage}}. Springer, \bibinfo{pages}{76--85}.
\newblock


\bibitem[Alsultan and Warwick(2013)]%
        {alsultan2013keystroke}
\bibfield{author}{\bibinfo{person}{Arwa Alsultan} {and} \bibinfo{person}{Kevin Warwick}.} \bibinfo{year}{2013}\natexlab{}.
\newblock \showarticletitle{Keystroke dynamics authentication: a survey of free-text methods}.
\newblock \bibinfo{journal}{\emph{International Journal of Computer Science Issues (IJCSI)}} \bibinfo{volume}{10}, \bibinfo{number}{4} (\bibinfo{year}{2013}), \bibinfo{pages}{1}.
\newblock


\bibitem[Antal and Denes-Fazakas(2019)]%
        {antal2019user}
\bibfield{author}{\bibinfo{person}{Margit Antal} {and} \bibinfo{person}{Lehel Denes-Fazakas}.} \bibinfo{year}{2019}\natexlab{}.
\newblock \showarticletitle{User Verification Based on Mouse Dynamics: a Comparison of Public Data Sets}. In \bibinfo{booktitle}{\emph{2019 IEEE 13th International Symposium on Applied Computational Intelligence and Informatics (SACI)}}. IEEE, \bibinfo{pages}{143--148}.
\newblock


\bibitem[Antal and Egyed-Zsigmond(2019)]%
        {antal2019intrusion}
\bibfield{author}{\bibinfo{person}{Margit Antal} {and} \bibinfo{person}{El{\"o}d Egyed-Zsigmond}.} \bibinfo{year}{2019}\natexlab{}.
\newblock \showarticletitle{Intrusion detection using mouse dynamics}.
\newblock \bibinfo{journal}{\emph{IET Biometrics}} \bibinfo{volume}{8}, \bibinfo{number}{5} (\bibinfo{year}{2019}), \bibinfo{pages}{285--294}.
\newblock


\bibitem[Antal and Fej{\'e}r(2020)]%
        {antal2020mouse}
\bibfield{author}{\bibinfo{person}{Margit Antal} {and} \bibinfo{person}{Norbert Fej{\'e}r}.} \bibinfo{year}{2020}\natexlab{}.
\newblock \showarticletitle{Mouse dynamics based user recognition using deep learning}.
\newblock \bibinfo{journal}{\emph{Acta Universitatis Sapientiae, Informatica}} \bibinfo{volume}{12}, \bibinfo{number}{1} (\bibinfo{year}{2020}), \bibinfo{pages}{39--50}.
\newblock


\bibitem[Antal et~al\mbox{.}(2021)]%
        {antal2021sapimouse}
\bibfield{author}{\bibinfo{person}{Margit Antal}, \bibinfo{person}{Norbert Fej{\'e}r}, {and} \bibinfo{person}{Krisztian Buza}.} \bibinfo{year}{2021}\natexlab{}.
\newblock \showarticletitle{SapiMouse: Mouse Dynamics-based User Authentication Using Deep Feature Learning}. In \bibinfo{booktitle}{\emph{2021 IEEE 15th International Symposium on Applied Computational Intelligence and Informatics (SACI)}}. IEEE, \bibinfo{pages}{61--66}.
\newblock


\bibitem[Atterer et~al\mbox{.}(2006)]%
        {atterer2006knowing}
\bibfield{author}{\bibinfo{person}{Richard Atterer}, \bibinfo{person}{Monika Wnuk}, {and} \bibinfo{person}{Albrecht Schmidt}.} \bibinfo{year}{2006}\natexlab{}.
\newblock \showarticletitle{Knowing the user's every move: user activity tracking for website usability evaluation and implicit interaction}. In \bibinfo{booktitle}{\emph{Proceedings of the 15th international conference on World Wide Web}}. \bibinfo{pages}{203--212}.
\newblock


\bibitem[Bailey et~al\mbox{.}(2014)]%
        {bailey2014user}
\bibfield{author}{\bibinfo{person}{Kyle~O Bailey}, \bibinfo{person}{James~S Okolica}, {and} \bibinfo{person}{Gilbert~L Peterson}.} \bibinfo{year}{2014}\natexlab{}.
\newblock \showarticletitle{User identification and authentication using multi-modal behavioral biometrics}.
\newblock \bibinfo{journal}{\emph{Computers \& Security}}  \bibinfo{volume}{43} (\bibinfo{year}{2014}), \bibinfo{pages}{77--89}.
\newblock


\bibitem[Balabit(2016)]%
        {balabit_2016}
\bibfield{author}{\bibinfo{person}{Balabit}.} \bibinfo{year}{2016}\natexlab{}.
\newblock \bibinfo{title}{balabit/mouse dynamics challenge}.
\newblock
\newblock
\urldef\tempurl%
\url{https://github.com/balabit/Mouse-Dynamics-Challenge}
\showURL{%
\tempurl}


\bibitem[Balaganesh and Sonia(2014)]%
        {balaganesh2014survey}
\bibfield{author}{\bibinfo{person}{PM Balaganesh} {and} \bibinfo{person}{A Sonia}.} \bibinfo{year}{2014}\natexlab{}.
\newblock \showarticletitle{A survey of authentication based on mouse behaviour}.
\newblock \bibinfo{journal}{\emph{international journal of advanced information science and technology}} \bibinfo{volume}{22}, \bibinfo{number}{22} (\bibinfo{year}{2014}).
\newblock


\bibitem[Banerjee and Woodard(2012)]%
        {banerjee2012biometric}
\bibfield{author}{\bibinfo{person}{Salil~P Banerjee} {and} \bibinfo{person}{Damon~L Woodard}.} \bibinfo{year}{2012}\natexlab{}.
\newblock \showarticletitle{Biometric authentication and identification using keystroke dynamics: A survey}.
\newblock \bibinfo{journal}{\emph{Journal of Pattern Recognition Research}} \bibinfo{volume}{7}, \bibinfo{number}{1} (\bibinfo{year}{2012}), \bibinfo{pages}{116--139}.
\newblock


\bibitem[Belman et~al\mbox{.}(2019)]%
        {belman2019insights}
\bibfield{author}{\bibinfo{person}{Amith~K Belman}, \bibinfo{person}{Li Wang}, \bibinfo{person}{SS Iyengar}, \bibinfo{person}{Pawel Sniatala}, \bibinfo{person}{Robert Wright}, \bibinfo{person}{Robert Dora}, \bibinfo{person}{Jacob Baldwin}, \bibinfo{person}{Zhanpeng Jin}, {and} \bibinfo{person}{Vir~V Phoha}.} \bibinfo{year}{2019}\natexlab{}.
\newblock \showarticletitle{Insights from BB-MAS--A Large Dataset for Typing, Gait and Swipes of the Same Person on Desktop, Tablet and Phone}.
\newblock \bibinfo{journal}{\emph{arXiv preprint arXiv:1912.02736}} (\bibinfo{year}{2019}).
\newblock


\bibitem[Bours and Fullu(2009)]%
        {bours2009login}
\bibfield{author}{\bibinfo{person}{Patrick Bours} {and} \bibinfo{person}{Christopher~Johnsrud Fullu}.} \bibinfo{year}{2009}\natexlab{}.
\newblock \showarticletitle{A login system using mouse dynamics}. In \bibinfo{booktitle}{\emph{2009 Fifth International Conference on Intelligent Information Hiding and Multimedia Signal Processing}}. IEEE, \bibinfo{pages}{1072--1077}.
\newblock


\bibitem[Bryan and Harter(1897)]%
        {bryan1897studies}
\bibfield{author}{\bibinfo{person}{William~Lowe Bryan} {and} \bibinfo{person}{Noble Harter}.} \bibinfo{year}{1897}\natexlab{}.
\newblock \showarticletitle{Studies in the physiology and psychology of the telegraphic language.}
\newblock \bibinfo{journal}{\emph{Psychological Review}} \bibinfo{volume}{4}, \bibinfo{number}{1} (\bibinfo{year}{1897}), \bibinfo{pages}{27}.
\newblock


\bibitem[Card et~al\mbox{.}(1978)]%
        {card1978evaluation}
\bibfield{author}{\bibinfo{person}{Stuart~K Card}, \bibinfo{person}{William~K English}, {and} \bibinfo{person}{Betty~J Burr}.} \bibinfo{year}{1978}\natexlab{}.
\newblock \showarticletitle{Evaluation of mouse, rate-controlled isometric joystick, step keys, and text keys for text selection on a CRT}.
\newblock \bibinfo{journal}{\emph{Ergonomics}} \bibinfo{volume}{21}, \bibinfo{number}{8} (\bibinfo{year}{1978}), \bibinfo{pages}{601--613}.
\newblock


\bibitem[Carneiro et~al\mbox{.}(2015)]%
        {carneiro2015using}
\bibfield{author}{\bibinfo{person}{Davide Carneiro}, \bibinfo{person}{Paulo Novais}, \bibinfo{person}{Jos{\'e}~Miguel P{\^e}go}, \bibinfo{person}{Nuno Sousa}, {and} \bibinfo{person}{Jos{\'e} Neves}.} \bibinfo{year}{2015}\natexlab{}.
\newblock \showarticletitle{Using mouse dynamics to assess stress during online exams}. In \bibinfo{booktitle}{\emph{International Conference on Hybrid Artificial Intelligence Systems}}. Springer, \bibinfo{pages}{345--356}.
\newblock


\bibitem[Carroll(1997)]%
        {carroll1997human}
\bibfield{author}{\bibinfo{person}{John~M Carroll}.} \bibinfo{year}{1997}\natexlab{}.
\newblock \showarticletitle{Human-computer interaction: psychology as a science of design}.
\newblock \bibinfo{journal}{\emph{Annual review of psychology}} \bibinfo{volume}{48}, \bibinfo{number}{1} (\bibinfo{year}{1997}), \bibinfo{pages}{61--83}.
\newblock


\bibitem[Case and Swanson(1998)]%
        {case1998constructing}
\bibfield{author}{\bibinfo{person}{Susan~M Case} {and} \bibinfo{person}{David~B Swanson}.} \bibinfo{year}{1998}\natexlab{}.
\newblock \bibinfo{booktitle}{\emph{Constructing written test questions for the basic and clinical sciences}}.
\newblock \bibinfo{publisher}{National Board of Medical Examiners Philadelphia, PA}.
\newblock


\bibitem[Chong et~al\mbox{.}(2019)]%
        {chong2019user}
\bibfield{author}{\bibinfo{person}{Penny Chong}, \bibinfo{person}{Yuval Elovici}, {and} \bibinfo{person}{Alexander Binder}.} \bibinfo{year}{2019}\natexlab{}.
\newblock \showarticletitle{User authentication based on mouse dynamics using deep neural networks: A comprehensive study}.
\newblock \bibinfo{journal}{\emph{IEEE Transactions on Information Forensics and Security}}  \bibinfo{volume}{15} (\bibinfo{year}{2019}), \bibinfo{pages}{1086--1101}.
\newblock


\bibitem[Chong et~al\mbox{.}(2018)]%
        {chong2018mouse}
\bibfield{author}{\bibinfo{person}{Penny Chong}, \bibinfo{person}{Yi~Xiang~Marcus Tan}, \bibinfo{person}{Juan Guarnizo}, \bibinfo{person}{Yuval Elovici}, {and} \bibinfo{person}{Alexander Binder}.} \bibinfo{year}{2018}\natexlab{}.
\newblock \showarticletitle{Mouse authentication without the temporal aspect--what does a 2d-cnn learn?}. In \bibinfo{booktitle}{\emph{2018 IEEE Security and Privacy Workshops (SPW)}}. IEEE, \bibinfo{pages}{15--21}.
\newblock


\bibitem[Ernsberger et~al\mbox{.}(2017)]%
        {ernsberger2017web}
\bibfield{author}{\bibinfo{person}{Dominik Ernsberger}, \bibinfo{person}{R~Adeyemi Ikuesan}, \bibinfo{person}{S~Hein Venter}, {and} \bibinfo{person}{Alf Zugenmaier}.} \bibinfo{year}{2017}\natexlab{}.
\newblock \showarticletitle{A web-based mouse dynamics visualization tool for user attribution in digital forensic readiness}. In \bibinfo{booktitle}{\emph{International Conference on Digital Forensics and Cyber Crime}}. Springer, \bibinfo{pages}{64--79}.
\newblock


\bibitem[et~al.(2007)]%
        {Dataset}
\bibfield{author}{\bibinfo{person}{Ahmed et al.}} \bibinfo{year}{2007}\natexlab{}.
\newblock \bibinfo{title}{ISOT Mouse Dynamics Dataset, University of Victoria}.
\newblock
\newblock
\urldef\tempurl%
\url{https://www.uvic.ca/engineering/ece/isot/datasets/behavioral-biometric/index.php}
\showURL{%
\tempurl}


\bibitem[European Standard EN 50133-1 Access control systems for use in~security applications({[n.\,d.]})]%
        {European_standard}
\bibfield{author}{\bibinfo{person}{Standard Number EN 50J33-J : J996IAJ: 2002 Technical Body CLClTC 79 European Committee for Electrotechnical Standardization (CENELEC)~2002 European Standard EN 50133-1 Access control systems for use in~security applications, Part 1: System~requirements}.} \bibinfo{year}{[n.\,d.]}\natexlab{}.
\newblock
\newblock
\urldef\tempurl%
\url{https://standards.iteh.ai/catalog/standards/clc/3f3cd487-5dcd-45be-a3ff-ae9ab5d69eac/en-50133-1-1996}
\showURL{%
\tempurl}


\bibitem[Everitt and McOwan(2003)]%
        {everitt2003java}
\bibfield{author}{\bibinfo{person}{Ross~AJ Everitt} {and} \bibinfo{person}{Peter~W McOwan}.} \bibinfo{year}{2003}\natexlab{}.
\newblock \showarticletitle{Java-based internet biometric authentication system}.
\newblock \bibinfo{journal}{\emph{IEEE Transactions on Pattern Analysis and Machine Intelligence}} \bibinfo{volume}{25}, \bibinfo{number}{9} (\bibinfo{year}{2003}), \bibinfo{pages}{1166--1172}.
\newblock


\bibitem[FBI(2017)]%
        {the_united_states_department_of_justice_2017}
\bibfield{author}{\bibinfo{person}{FBI}.} \bibinfo{year}{2017}\natexlab{}.
\newblock \bibinfo{title}{U.S. Charges Russian FSB Officers and Their Criminal Conspirators for Hacking Yahoo and Millions of Email Accounts}.
\newblock
\newblock
\urldef\tempurl%
\url{https://www.justice.gov/opa/pr/us-charges-russian-fsb-officers-and-their-criminal-conspirators-hacking-yahoo-and-millions}
\showURL{%
\tempurl}


\bibitem[FBI(2020)]%
        {the_united_states_department_of_justice_2020}
\bibfield{author}{\bibinfo{person}{FBI}.} \bibinfo{year}{2020}\natexlab{}.
\newblock \bibinfo{title}{Chinese Military Personnel Charged with Computer Fraud, Economic Espionage and Wire Fraud for Hacking into Credit Reporting Agency Equifax}.
\newblock
\newblock
\urldef\tempurl%
\url{https://www.justice.gov/opa/pr/chinese-military-personnel-charged-computer-fraud-economic-espionage-and-wire-fraud-hacking}
\showURL{%
\tempurl}


\bibitem[Feher et~al\mbox{.}(2012)]%
        {feher2012user}
\bibfield{author}{\bibinfo{person}{Clint Feher}, \bibinfo{person}{Yuval Elovici}, \bibinfo{person}{Robert Moskovitch}, \bibinfo{person}{Lior Rokach}, {and} \bibinfo{person}{Alon Schclar}.} \bibinfo{year}{2012}\natexlab{}.
\newblock \showarticletitle{User identity verification via mouse dynamics}.
\newblock \bibinfo{journal}{\emph{Information Sciences}}  \bibinfo{volume}{201} (\bibinfo{year}{2012}), \bibinfo{pages}{19--36}.
\newblock


\bibitem[Fitts(1954)]%
        {fitts1954information}
\bibfield{author}{\bibinfo{person}{Paul~M Fitts}.} \bibinfo{year}{1954}\natexlab{}.
\newblock \showarticletitle{The information capacity of the human motor system in controlling the amplitude of movement}.
\newblock \bibinfo{journal}{\emph{Journal of experimental psychology}} \bibinfo{volume}{47}, \bibinfo{number}{6} (\bibinfo{year}{1954}), \bibinfo{pages}{381}.
\newblock


\bibitem[Fu et~al\mbox{.}(2022)]%
        {fu2022}
\bibfield{author}{\bibinfo{person}{Shen Fu}, \bibinfo{person}{Dong Qin}, \bibinfo{person}{George Amariucai}, \bibinfo{person}{Daji Qiao}, \bibinfo{person}{Yong Guan}, {and} \bibinfo{person}{Ann Smiley}.} \bibinfo{year}{2022}\natexlab{}.
\newblock \showarticletitle{Artificial Intelligence Meets Kinesthetic Intelligence: Mouse-based User Authentication based on Hybrid Human-Machine Learning}. In \bibinfo{booktitle}{\emph{Proceedings of the 2022 ACM on Asia Conference on Computer and Communications Security}}. \bibinfo{pages}{1034--1048}.
\newblock


\bibitem[Gamboa and Fred(2003)]%
        {gamboa2003identity}
\bibfield{author}{\bibinfo{person}{Hugo Gamboa} {and} \bibinfo{person}{Ana~LN Fred}.} \bibinfo{year}{2003}\natexlab{}.
\newblock \showarticletitle{An Identity Authentication System Based On Human Computer Interaction Behaviour.}. In \bibinfo{booktitle}{\emph{PRIS}}. \bibinfo{pages}{46--55}.
\newblock


\bibitem[Gao et~al\mbox{.}(2020)]%
        {gao2020continuous}
\bibfield{author}{\bibinfo{person}{Lifang Gao}, \bibinfo{person}{Yangyang Lian}, \bibinfo{person}{Huifeng Yang}, \bibinfo{person}{Rui Xin}, \bibinfo{person}{Zhuozhi Yu}, \bibinfo{person}{Wenwei Chen}, \bibinfo{person}{Wei Liu}, \bibinfo{person}{Yefeng Zhang}, \bibinfo{person}{Yukun Zhu}, \bibinfo{person}{Siya Xu}, {et~al\mbox{.}}} \bibinfo{year}{2020}\natexlab{}.
\newblock \showarticletitle{Continuous authentication of mouse dynamics based on decision level fusion}. In \bibinfo{booktitle}{\emph{2020 International Wireless Communications and Mobile Computing (IWCMC)}}. IEEE, \bibinfo{pages}{210--214}.
\newblock


\bibitem[Goecks and Shavlik(1999)]%
        {goecks1999automatically}
\bibfield{author}{\bibinfo{person}{Jeremy Goecks} {and} \bibinfo{person}{Jude Shavlik}.} \bibinfo{year}{1999}\natexlab{}.
\newblock \showarticletitle{Automatically labeling web pages based on normal user actions}. In \bibinfo{booktitle}{\emph{Procedings of the IJCAI Workshop on Machine Learning for Information Filtering}}.
\newblock


\bibitem[Gray and Boehm-Davis(2000)]%
        {gray2000milliseconds}
\bibfield{author}{\bibinfo{person}{Wayne~D Gray} {and} \bibinfo{person}{Deborah~A Boehm-Davis}.} \bibinfo{year}{2000}\natexlab{}.
\newblock \showarticletitle{Milliseconds matter: An introduction to microstrategies and to their use in describing and predicting interactive behavior.}
\newblock \bibinfo{journal}{\emph{Journal of experimental psychology: applied}} \bibinfo{volume}{6}, \bibinfo{number}{4} (\bibinfo{year}{2000}), \bibinfo{pages}{322}.
\newblock


\bibitem[Gray et~al\mbox{.}(1993)]%
        {gray1993project}
\bibfield{author}{\bibinfo{person}{Wayne~D Gray}, \bibinfo{person}{Bonnie~E John}, {and} \bibinfo{person}{Michael~E Atwood}.} \bibinfo{year}{1993}\natexlab{}.
\newblock \showarticletitle{Project Ernestine: Validating a GOMS analysis for predicting and explaining real-world task performance}.
\newblock \bibinfo{journal}{\emph{Human-computer interaction}} \bibinfo{volume}{8}, \bibinfo{number}{3} (\bibinfo{year}{1993}), \bibinfo{pages}{237--309}.
\newblock


\bibitem[Group(2005)]%
        {international_biometric_group_2005}
\bibfield{author}{\bibinfo{person}{International~Biometric Group}.} \bibinfo{year}{2005}\natexlab{}.
\newblock \bibinfo{title}{Independent Testing of Iris Recognition Technology}.
\newblock
\newblock
\urldef\tempurl%
\url{https://www.hsdl.org/?view&did=464567}
\showURL{%
\tempurl}


\bibitem[Gupta(2020)]%
        {gupta_2020}
\bibfield{author}{\bibinfo{person}{Mehul Gupta}.} \bibinfo{year}{2020}\natexlab{}.
\newblock \bibinfo{title}{Dimension Reduction using Isomap}.
\newblock
\newblock
\urldef\tempurl%
\url{https://medium.com/data-science-in-your-pocket/dimension-reduction-using-isomap-72ead0411dec}
\showURL{%
\tempurl}


\bibitem[Hamdy and Traor{\'e}(2011)]%
        {hamdy2011}
\bibfield{author}{\bibinfo{person}{Omar Hamdy} {and} \bibinfo{person}{Issa Traor{\'e}}.} \bibinfo{year}{2011}\natexlab{}.
\newblock \showarticletitle{Homogeneous physio-behavioral visual and mouse-based biometric}.
\newblock \bibinfo{journal}{\emph{ACM Transactions on Computer-Human Interaction (TOCHI)}} \bibinfo{volume}{18}, \bibinfo{number}{3} (\bibinfo{year}{2011}), \bibinfo{pages}{1--30}.
\newblock


\bibitem[Harilal et~al\mbox{.}(2017)]%
        {harilal2017twos}
\bibfield{author}{\bibinfo{person}{Athul Harilal}, \bibinfo{person}{Flavio Toffalini}, \bibinfo{person}{John Castellanos}, \bibinfo{person}{Juan Guarnizo}, \bibinfo{person}{Ivan Homoliak}, {and} \bibinfo{person}{Mart{\'\i}n Ochoa}.} \bibinfo{year}{2017}\natexlab{}.
\newblock \showarticletitle{Twos: A dataset of malicious insider threat behavior based on a gamified competition}. In \bibinfo{booktitle}{\emph{Proceedings of the 2017 International Workshop on Managing Insider Security Threats}}. \bibinfo{pages}{45--56}.
\newblock


\bibitem[Hashia et~al\mbox{.}(2005)]%
        {hashia2005using}
\bibfield{author}{\bibinfo{person}{Shivani Hashia}, \bibinfo{person}{Chris Pollett}, {and} \bibinfo{person}{Mark Stamp}.} \bibinfo{year}{2005}\natexlab{}.
\newblock \showarticletitle{On using mouse movements as a biometric}. In \bibinfo{booktitle}{\emph{Proceeding in the International Conference on Computer Science and its Applications}}, Vol.~\bibinfo{volume}{1}. The International Conference on Computer Science and its Applications (ICCSA~…, \bibinfo{pages}{5}.
\newblock


\bibitem[Hassan~Hosny et~al\mbox{.}(2022)]%
        {hassan2022intelligent}
\bibfield{author}{\bibinfo{person}{Hadeer~A Hassan~Hosny}, \bibinfo{person}{Abdulrahman~A Ibrahim}, \bibinfo{person}{Mahmoud~M Elmesalawy}, {and} \bibinfo{person}{Ahmed~M Abd El-Haleem}.} \bibinfo{year}{2022}\natexlab{}.
\newblock \showarticletitle{An Intelligent Approach for Fair Assessment of Online Laboratory Examinations in Laboratory Learning Systems Based on Student’s Mouse Interaction Behavior}.
\newblock \bibinfo{journal}{\emph{Applied Sciences}} \bibinfo{volume}{12}, \bibinfo{number}{22} (\bibinfo{year}{2022}), \bibinfo{pages}{11416}.
\newblock


\bibitem[Hick(1952)]%
        {hick1952rate}
\bibfield{author}{\bibinfo{person}{William~E Hick}.} \bibinfo{year}{1952}\natexlab{}.
\newblock \showarticletitle{On the rate of gain of information}.
\newblock \bibinfo{journal}{\emph{Quarterly Journal of experimental psychology}} \bibinfo{volume}{4}, \bibinfo{number}{1} (\bibinfo{year}{1952}), \bibinfo{pages}{11--26}.
\newblock


\bibitem[Hickeys(2012)]%
        {hickeys_2012}
\bibfield{author}{\bibinfo{person}{Hickeys}.} \bibinfo{year}{2012}\natexlab{}.
\newblock \bibinfo{title}{Hooks Overview - Win32 apps}.
\newblock
\newblock
\urldef\tempurl%
\url{https://docs.microsoft.com/en-us/windows/win32/winmsg/about-hooks}
\showURL{%
\tempurl}


\bibitem[Hinbarji et~al\mbox{.}(2015)]%
        {hinbarji2015dynamic}
\bibfield{author}{\bibinfo{person}{Zaher Hinbarji}, \bibinfo{person}{Rami Albatal}, {and} \bibinfo{person}{Cathal Gurrin}.} \bibinfo{year}{2015}\natexlab{}.
\newblock \showarticletitle{Dynamic user authentication based on mouse movements curves}. In \bibinfo{booktitle}{\emph{International Conference on Multimedia Modeling}}. Springer, \bibinfo{pages}{111--122}.
\newblock


\bibitem[Hoffmann(1991)]%
        {hoffmann1991}
\bibfield{author}{\bibinfo{person}{Errol~R Hoffmann}.} \bibinfo{year}{1991}\natexlab{}.
\newblock \showarticletitle{Capture of moving targets: a modification of Fitts' Law}.
\newblock \bibinfo{journal}{\emph{Ergonomics}} \bibinfo{volume}{34}, \bibinfo{number}{2} (\bibinfo{year}{1991}), \bibinfo{pages}{211--220}.
\newblock


\bibitem[Hu et~al\mbox{.}(2017)]%
        {hu2017deceive}
\bibfield{author}{\bibinfo{person}{Shujie Hu}, \bibinfo{person}{Jun Bai}, \bibinfo{person}{Hongri Liu}, \bibinfo{person}{Chao Wang}, {and} \bibinfo{person}{Bailing Wang}.} \bibinfo{year}{2017}\natexlab{}.
\newblock \showarticletitle{Deceive mouse-dynamics-based authentication model via movement simulation}. In \bibinfo{booktitle}{\emph{2017 10th International Symposium on Computational Intelligence and Design (ISCID)}}, Vol.~\bibinfo{volume}{1}. IEEE, \bibinfo{pages}{482--485}.
\newblock


\bibitem[Hu et~al\mbox{.}(2019)]%
        {hu2019insider}
\bibfield{author}{\bibinfo{person}{Teng Hu}, \bibinfo{person}{Weina Niu}, \bibinfo{person}{Xiaosong Zhang}, \bibinfo{person}{Xiaolei Liu}, \bibinfo{person}{Jiazhong Lu}, {and} \bibinfo{person}{Yuan Liu}.} \bibinfo{year}{2019}\natexlab{}.
\newblock \showarticletitle{An insider threat detection approach based on mouse dynamics and deep learning}.
\newblock \bibinfo{journal}{\emph{Security and Communication Networks}}  \bibinfo{volume}{2019} (\bibinfo{year}{2019}).
\newblock


\bibitem[Huang et~al\mbox{.}(2020)]%
        {huang2020}
\bibfield{author}{\bibinfo{person}{Jin Huang}, \bibinfo{person}{Feng Tian}, \bibinfo{person}{Xiangmin Fan}, \bibinfo{person}{Huawei Tu}, \bibinfo{person}{Hao Zhang}, \bibinfo{person}{Xiaolan Peng}, {and} \bibinfo{person}{Hongan Wang}.} \bibinfo{year}{2020}\natexlab{}.
\newblock \showarticletitle{Modeling the endpoint uncertainty in crossing-based moving target selection}. In \bibinfo{booktitle}{\emph{Proceedings of the 2020 CHI Conference on Human Factors in Computing Systems}}. \bibinfo{pages}{1--12}.
\newblock


\bibitem[Huang et~al\mbox{.}(2018)]%
        {huang2018}
\bibfield{author}{\bibinfo{person}{Jin Huang}, \bibinfo{person}{Feng Tian}, \bibinfo{person}{Xiangmin Fan}, \bibinfo{person}{Xiaolong Zhang}, {and} \bibinfo{person}{Shumin Zhai}.} \bibinfo{year}{2018}\natexlab{}.
\newblock \showarticletitle{Understanding the uncertainty in 1D unidirectional moving target selection}. In \bibinfo{booktitle}{\emph{Proceedings of the 2018 CHI conference on human factors in computing systems}}. \bibinfo{pages}{1--12}.
\newblock


\bibitem[Huang et~al\mbox{.}(2019)]%
        {huang2019}
\bibfield{author}{\bibinfo{person}{Jin Huang}, \bibinfo{person}{Feng Tian}, \bibinfo{person}{Nianlong Li}, {and} \bibinfo{person}{Xiangmin Fan}.} \bibinfo{year}{2019}\natexlab{}.
\newblock \showarticletitle{Modeling the uncertainty in 2D moving target selection}. In \bibinfo{booktitle}{\emph{Proceedings of the 32nd annual ACM symposium on user interface software and technology}}. \bibinfo{pages}{1031--1043}.
\newblock


\bibitem[Imsand(2008)]%
        {imsand2008applications}
\bibfield{author}{\bibinfo{person}{Eric~Shaun Imsand}.} \bibinfo{year}{2008}\natexlab{}.
\newblock \bibinfo{booktitle}{\emph{Applications of GUI usage analysis}}.
\newblock \bibinfo{publisher}{Auburn University}.
\newblock


\bibitem[Jaadi({[n.\,d.]})]%
        {jaadi}
\bibfield{author}{\bibinfo{person}{Zakaria Jaadi}.} \bibinfo{year}{[n.\,d.]}\natexlab{}.
\newblock \bibinfo{title}{A Step-by-Step Explanation of Principal Component Analysis (PCA)}.
\newblock
\newblock
\urldef\tempurl%
\url{https://builtin.com/data-science/step-step-explanation-principal-component-analysis}
\showURL{%
\tempurl}


\bibitem[Jagacinski et~al\mbox{.}(1980)]%
        {jagacinski1980}
\bibfield{author}{\bibinfo{person}{Richard~J Jagacinski}, \bibinfo{person}{Daniel~W Repperger}, \bibinfo{person}{Sharon~L Ward}, {and} \bibinfo{person}{Martin~S Moran}.} \bibinfo{year}{1980}\natexlab{}.
\newblock \showarticletitle{A test of Fitts' law with moving targets}.
\newblock \bibinfo{journal}{\emph{Human Factors}} \bibinfo{volume}{22}, \bibinfo{number}{2} (\bibinfo{year}{1980}), \bibinfo{pages}{225--233}.
\newblock


\bibitem[Jain et~al\mbox{.}(2006)]%
        {jain2006biometrics}
\bibfield{author}{\bibinfo{person}{Anil~K Jain}, \bibinfo{person}{Arun Ross}, {and} \bibinfo{person}{Sharath Pankanti}.} \bibinfo{year}{2006}\natexlab{}.
\newblock \showarticletitle{Biometrics: a tool for information security}.
\newblock \bibinfo{journal}{\emph{IEEE transactions on information forensics and security}} \bibinfo{volume}{1}, \bibinfo{number}{2} (\bibinfo{year}{2006}), \bibinfo{pages}{125--143}.
\newblock


\bibitem[John and Kieras(1996)]%
        {john1996goms}
\bibfield{author}{\bibinfo{person}{Bonnie~E John} {and} \bibinfo{person}{David~E Kieras}.} \bibinfo{year}{1996}\natexlab{}.
\newblock \showarticletitle{The GOMS family of user interface analysis techniques: Comparison and contrast}.
\newblock \bibinfo{journal}{\emph{ACM Transactions on Computer-Human Interaction (TOCHI)}} \bibinfo{volume}{3}, \bibinfo{number}{4} (\bibinfo{year}{1996}), \bibinfo{pages}{320--351}.
\newblock


\bibitem[Jorgensen and Yu(2011)]%
        {jorgensen2011mouse}
\bibfield{author}{\bibinfo{person}{Zach Jorgensen} {and} \bibinfo{person}{Ting Yu}.} \bibinfo{year}{2011}\natexlab{}.
\newblock \showarticletitle{On mouse dynamics as a behavioral biometric for authentication}. In \bibinfo{booktitle}{\emph{Proceedings of the 6th ACM Symposium on Information, Computer and Communications Security}}. \bibinfo{pages}{476--482}.
\newblock


\bibitem[Kaixin et~al\mbox{.}(2017)]%
        {kaixin2017user}
\bibfield{author}{\bibinfo{person}{Wang Kaixin}, \bibinfo{person}{Liu Hongri}, \bibinfo{person}{Wang Bailing}, \bibinfo{person}{Hu Shujie}, {and} \bibinfo{person}{Song jia}.} \bibinfo{year}{2017}\natexlab{}.
\newblock \showarticletitle{A User Authentication and Identification Model Based on Mouse Dynamics}. In \bibinfo{booktitle}{\emph{Proceedings of the 6th International Conference on Information Engineering}}. \bibinfo{pages}{1--6}.
\newblock


\bibitem[Kaminsky et~al\mbox{.}(2008)]%
        {kaminsky2008identifying}
\bibfield{author}{\bibinfo{person}{Ryan Kaminsky}, \bibinfo{person}{Miro Enev}, {and} \bibinfo{person}{Erik Andersen}.} \bibinfo{year}{2008}\natexlab{}.
\newblock \showarticletitle{Identifying game players with mouse biometrics}.
\newblock \bibinfo{journal}{\emph{University of Washington. Technical Report}} (\bibinfo{year}{2008}).
\newblock


\bibitem[Kang and Kim(2023)]%
        {kang2023user}
\bibfield{author}{\bibinfo{person}{Shin~Jin Kang} {and} \bibinfo{person}{Soo~Kyun Kim}.} \bibinfo{year}{2023}\natexlab{}.
\newblock \showarticletitle{User Interface-Based Repeated Sequence Detection Method for Authentication}.
\newblock \bibinfo{journal}{\emph{Intelligent Automation and Soft Computing}} \bibinfo{volume}{35}, \bibinfo{number}{3} (\bibinfo{year}{2023}), \bibinfo{pages}{2573--2588}.
\newblock


\bibitem[Khan et~al\mbox{.}({[n.\,d.]})]%
        {khanmouse}
\bibfield{author}{\bibinfo{person}{Anam Khan}, \bibinfo{person}{Suhail~Javed Quraishi}, {and} \bibinfo{person}{Sarabjeet~Singh Bedi}.} \bibinfo{year}{[n.\,d.]}\natexlab{}.
\newblock \showarticletitle{Mouse Dynamics as Continuous User Authentication Tool‖}.
\newblock \bibinfo{journal}{\emph{International Journal of Recent Technology and Engineering (IJRTE), ISSN}} (\bibinfo{year}{[n.\,d.]}), \bibinfo{pages}{2277--3878}.
\newblock


\bibitem[Khan et~al\mbox{.}(2021)]%
        {khan2021authenticating}
\bibfield{author}{\bibinfo{person}{Simon Khan}, \bibinfo{person}{Cooper Fraser}, \bibinfo{person}{Daqing Hou}, \bibinfo{person}{Mahesh Banavar}, {and} \bibinfo{person}{Stephanie Schuckers}.} \bibinfo{year}{2021}\natexlab{}.
\newblock \showarticletitle{Authenticating Facebook Users Based on Widget Interaction Behavior}. In \bibinfo{booktitle}{\emph{2021 IEEE 18th Annual Consumer Communications \& Networking Conference (CCNC)}}. IEEE, \bibinfo{pages}{1--8}.
\newblock


\bibitem[Kukreja et~al\mbox{.}(2006)]%
        {kukreja2006rui}
\bibfield{author}{\bibinfo{person}{Urmila Kukreja}, \bibinfo{person}{William~E Stevenson}, {and} \bibinfo{person}{Frank~E Ritter}.} \bibinfo{year}{2006}\natexlab{}.
\newblock \showarticletitle{RUI: Recording user input from interfaces under Windows and Mac OS X}.
\newblock \bibinfo{journal}{\emph{Behavior Research Methods}} \bibinfo{volume}{38}, \bibinfo{number}{4} (\bibinfo{year}{2006}), \bibinfo{pages}{656--659}.
\newblock


\bibitem[Kumar and Tomkins(2010)]%
        {kumar2010characterization}
\bibfield{author}{\bibinfo{person}{Ravi Kumar} {and} \bibinfo{person}{Andrew Tomkins}.} \bibinfo{year}{2010}\natexlab{}.
\newblock \showarticletitle{A characterization of online browsing behavior}. In \bibinfo{booktitle}{\emph{Proceedings of the 19th international conference on World wide web}}. \bibinfo{pages}{561--570}.
\newblock


\bibitem[Lane and Brodley(1997)]%
        {lane1997application}
\bibfield{author}{\bibinfo{person}{Terran Lane} {and} \bibinfo{person}{Carla~E Brodley}.} \bibinfo{year}{1997}\natexlab{}.
\newblock \showarticletitle{An application of machine learning to anomaly detection}. In \bibinfo{booktitle}{\emph{Proceedings of the 20th National Information Systems Security Conference}}, Vol.~\bibinfo{volume}{377}. Baltimore, USA, \bibinfo{pages}{366--380}.
\newblock


\bibitem[Lane and Brodley(1998)]%
        {lane1998approaches}
\bibfield{author}{\bibinfo{person}{Terran Lane} {and} \bibinfo{person}{Carla~E Brodley}.} \bibinfo{year}{1998}\natexlab{}.
\newblock \showarticletitle{Approaches to Online Learning and Concept Drift for User Identification in Computer Security.}. In \bibinfo{booktitle}{\emph{KDD}}. \bibinfo{pages}{259--263}.
\newblock


\bibitem[Lane and Brodley(1999)]%
        {lane1999temporal}
\bibfield{author}{\bibinfo{person}{Terran Lane} {and} \bibinfo{person}{Carla~E Brodley}.} \bibinfo{year}{1999}\natexlab{}.
\newblock \showarticletitle{Temporal sequence learning and data reduction for anomaly detection}.
\newblock \bibinfo{journal}{\emph{ACM Transactions on Information and System Security (TISSEC)}} \bibinfo{volume}{2}, \bibinfo{number}{3} (\bibinfo{year}{1999}), \bibinfo{pages}{295--331}.
\newblock


\bibitem[Lane et~al\mbox{.}(1997)]%
        {lane1997sequence}
\bibfield{author}{\bibinfo{person}{Terran Lane}, \bibinfo{person}{Carla~E Brodley}, {et~al\mbox{.}}} \bibinfo{year}{1997}\natexlab{}.
\newblock \showarticletitle{Sequence matching and learning in anomaly detection for computer security}. In \bibinfo{booktitle}{\emph{AAAI Workshop: AI Approaches to Fraud Detection and Risk Management}}. Providence, Rhode Island, \bibinfo{pages}{43--49}.
\newblock


\bibitem[Lee et~al\mbox{.}(2018)]%
        {lee2018}
\bibfield{author}{\bibinfo{person}{Byungjoo Lee}, \bibinfo{person}{Sunjun Kim}, \bibinfo{person}{Antti Oulasvirta}, \bibinfo{person}{Jong-In Lee}, {and} \bibinfo{person}{Eunji Park}.} \bibinfo{year}{2018}\natexlab{}.
\newblock \showarticletitle{Moving target selection: A cue integration model}. In \bibinfo{booktitle}{\emph{Proceedings of the 2018 CHI Conference on Human Factors in Computing Systems}}. \bibinfo{pages}{1--12}.
\newblock


\bibitem[Lee and Oulasvirta(2016)]%
        {lee2016}
\bibfield{author}{\bibinfo{person}{Byungjoo Lee} {and} \bibinfo{person}{Antti Oulasvirta}.} \bibinfo{year}{2016}\natexlab{}.
\newblock \showarticletitle{Modelling error rates in temporal pointing}. In \bibinfo{booktitle}{\emph{Proceedings of the 2016 CHI Conference on Human Factors in Computing Systems}}. \bibinfo{pages}{1857--1868}.
\newblock


\bibitem[Levy(2023)]%
        {levy}
\bibfield{author}{\bibinfo{person}{D Levy}.} \bibinfo{year}{2023}\natexlab{}.
\newblock \bibinfo{title}{Numerical Differentiation}.
\newblock
\newblock
\urldef\tempurl%
\url{http://www2.math.umd.edu/~dlevy/classes/amsc466/lecture-notes/differentiation-chap.pdf}
\showURL{%
\tempurl}


\bibitem[Li et~al\mbox{.}(2016)]%
        {li2016}
\bibfield{author}{\bibinfo{person}{Jiajia Li}, \bibinfo{person}{Grace Ngai}, \bibinfo{person}{Hong~Va Leong}, {and} \bibinfo{person}{Stephen~CF Chan}.} \bibinfo{year}{2016}\natexlab{}.
\newblock \showarticletitle{Multimodal human attention detection for reading from facial expression, eye gaze, and mouse dynamics}.
\newblock \bibinfo{journal}{\emph{ACM SIGAPP Applied Computing Review}} \bibinfo{volume}{16}, \bibinfo{number}{3} (\bibinfo{year}{2016}), \bibinfo{pages}{37--49}.
\newblock


\bibitem[L{\'o}pez et~al\mbox{.}(2023)]%
        {lopez2023adversarial}
\bibfield{author}{\bibinfo{person}{Christian L{\'o}pez}, \bibinfo{person}{Jes{\'u}s Solano}, \bibinfo{person}{Esteban Rivera}, \bibinfo{person}{Lizzy Tengana}, \bibinfo{person}{Johana Florez-Lozano}, \bibinfo{person}{Alejandra Castelblanco}, {and} \bibinfo{person}{Mart{\'\i}n Ochoa}.} \bibinfo{year}{2023}\natexlab{}.
\newblock \showarticletitle{Adversarial attacks against mouse-and keyboard-based biometric authentication: black-box versus domain-specific techniques}.
\newblock \bibinfo{journal}{\emph{International Journal of Information Security}} (\bibinfo{year}{2023}), \bibinfo{pages}{1--21}.
\newblock


\bibitem[Ma et~al\mbox{.}(2016)]%
        {ma2016kind}
\bibfield{author}{\bibinfo{person}{Lei Ma}, \bibinfo{person}{Chungang Yan}, \bibinfo{person}{Peihai Zhao}, {and} \bibinfo{person}{Mimi Wang}.} \bibinfo{year}{2016}\natexlab{}.
\newblock \showarticletitle{A kind of mouse behavior authentication method on dynamic soft keyboard}. In \bibinfo{booktitle}{\emph{2016 IEEE International Conference on Systems, Man, and Cybernetics (SMC)}}. IEEE, \bibinfo{pages}{000211--000216}.
\newblock


\bibitem[MacKenzie(1992)]%
        {mackenzie1992}
\bibfield{author}{\bibinfo{person}{I~Scott MacKenzie}.} \bibinfo{year}{1992}\natexlab{}.
\newblock \showarticletitle{Fitts' law as a research and design tool in human-computer interaction}.
\newblock \bibinfo{journal}{\emph{Human-computer interaction}} \bibinfo{volume}{7}, \bibinfo{number}{1} (\bibinfo{year}{1992}), \bibinfo{pages}{91--139}.
\newblock


\bibitem[Math24(2021)]%
        {math24_2021}
\bibfield{author}{\bibinfo{person}{Math24}.} \bibinfo{year}{2021}\natexlab{}.
\newblock \bibinfo{title}{Curvature and Radius of Curvature}.
\newblock
\newblock
\urldef\tempurl%
\url{https://www.math24.net/curvature-radius}
\showURL{%
\tempurl}


\bibitem[Mathworks(2019)]%
        {Differentiation}
\bibfield{author}{\bibinfo{person}{Mathworks}.} \bibinfo{year}{2019}\natexlab{}.
\newblock \bibinfo{title}{How to calculate the second and third numerical derivative of one variable f(x)}.
\newblock
\newblock
\urldef\tempurl%
\url{https://www.mathworks.com/matlabcentral/answers/496527-how-calculate-the-second-and-third-numerical-derivative-of-one-variable-f-x}
\showURL{%
\tempurl}


\bibitem[Maxion and Townsend(2002)]%
        {maxion2002masquerade}
\bibfield{author}{\bibinfo{person}{Roy~A Maxion} {and} \bibinfo{person}{Tahlia~N Townsend}.} \bibinfo{year}{2002}\natexlab{}.
\newblock \showarticletitle{Masquerade detection using truncated command lines}. In \bibinfo{booktitle}{\emph{Proceedings international conference on dependable systems and networks}}. IEEE, \bibinfo{pages}{219--228}.
\newblock


\bibitem[Mika et~al\mbox{.}(1998)]%
        {mika1998kernel}
\bibfield{author}{\bibinfo{person}{Sebastian Mika}, \bibinfo{person}{Bernhard Sch{\"o}lkopf}, \bibinfo{person}{Alexander~J Smola}, \bibinfo{person}{Klaus-Robert M{\"u}ller}, \bibinfo{person}{Matthias Scholz}, {and} \bibinfo{person}{Gunnar R{\"a}tsch}.} \bibinfo{year}{1998}\natexlab{}.
\newblock \showarticletitle{Kernel PCA and De-noising in feature spaces.}. In \bibinfo{booktitle}{\emph{NIPS}}, Vol.~\bibinfo{volume}{11}. \bibinfo{pages}{536--542}.
\newblock


\bibitem[MIT(2021)]%
        {Trajectory_Calculation}
\bibfield{author}{\bibinfo{person}{MIT}.} \bibinfo{year}{2021}\natexlab{}.
\newblock \bibinfo{title}{Lecture Notes}.
\newblock
\newblock
\urldef\tempurl%
\url{http://web.mit.edu/16.unified/www/FALL/systems/Lab_Notes/traj.pdf}
\showURL{%
\tempurl}


\bibitem[Mondal and Bours(2013)]%
        {mondal2013continuous}
\bibfield{author}{\bibinfo{person}{Soumik Mondal} {and} \bibinfo{person}{Patrick Bours}.} \bibinfo{year}{2013}\natexlab{}.
\newblock \showarticletitle{Continuous authentication using mouse dynamics}. In \bibinfo{booktitle}{\emph{2013 International Conference of the BIOSIG Special Interest Group (BIOSIG)}}. IEEE, \bibinfo{pages}{1--12}.
\newblock


\bibitem[Mondal and Bours(2017)]%
        {mondal2017study}
\bibfield{author}{\bibinfo{person}{Soumik Mondal} {and} \bibinfo{person}{Patrick Bours}.} \bibinfo{year}{2017}\natexlab{}.
\newblock \showarticletitle{A study on continuous authentication using a combination of keystroke and mouse biometrics}.
\newblock \bibinfo{journal}{\emph{Neurocomputing}}  \bibinfo{volume}{230} (\bibinfo{year}{2017}), \bibinfo{pages}{1--22}.
\newblock


\bibitem[Nakkabi et~al\mbox{.}(2010)]%
        {nakkabi2010improving}
\bibfield{author}{\bibinfo{person}{Youssef Nakkabi}, \bibinfo{person}{Issa Traor{\'e}}, {and} \bibinfo{person}{Ahmed Awad~E Ahmed}.} \bibinfo{year}{2010}\natexlab{}.
\newblock \showarticletitle{Improving mouse dynamics biometric performance using variance reduction via extractors with separate features}.
\newblock \bibinfo{journal}{\emph{IEEE Transactions on Systems, Man, and Cybernetics-Part A: Systems and Humans}} \bibinfo{volume}{40}, \bibinfo{number}{6} (\bibinfo{year}{2010}), \bibinfo{pages}{1345--1353}.
\newblock


\bibitem[Obaidat and Macchairolo(1994)]%
        {obaidat1994multilayer}
\bibfield{author}{\bibinfo{person}{MS Obaidat} {and} \bibinfo{person}{DT Macchairolo}.} \bibinfo{year}{1994}\natexlab{}.
\newblock \showarticletitle{A multilayer neural network system for computer access security}.
\newblock \bibinfo{journal}{\emph{IEEE Transactions on Systems, Man, and Cybernetics}} \bibinfo{volume}{24}, \bibinfo{number}{5} (\bibinfo{year}{1994}), \bibinfo{pages}{806--813}.
\newblock


\bibitem[Obaidat and Macchiarolo(1993)]%
        {obaidat1993online}
\bibfield{author}{\bibinfo{person}{Mohammad~S Obaidat} {and} \bibinfo{person}{David~T Macchiarolo}.} \bibinfo{year}{1993}\natexlab{}.
\newblock \showarticletitle{An online neural network system for computer access security}.
\newblock \bibinfo{journal}{\emph{IEEE Transactions on Industrial electronics}} \bibinfo{volume}{40}, \bibinfo{number}{2} (\bibinfo{year}{1993}), \bibinfo{pages}{235--242}.
\newblock


\bibitem[Olson and Olson(1995)]%
        {olson1995growth}
\bibfield{author}{\bibinfo{person}{Judith~Reitman Olson} {and} \bibinfo{person}{Gary~M Olson}.} \bibinfo{year}{1995}\natexlab{}.
\newblock \showarticletitle{The growth of cognitive modeling in human-computer interaction since GOMS}.
\newblock In \bibinfo{booktitle}{\emph{Readings in Human--Computer Interaction}}. \bibinfo{publisher}{Elsevier}, \bibinfo{pages}{603--625}.
\newblock


\bibitem[Orebaugh and Allnutt(2009)]%
        {orebaugh2009classification}
\bibfield{author}{\bibinfo{person}{Angela Orebaugh} {and} \bibinfo{person}{Jeremy Allnutt}.} \bibinfo{year}{2009}\natexlab{}.
\newblock \showarticletitle{Classification of instant messaging communications for forensics analysis}.
\newblock \bibinfo{journal}{\emph{The International Journal of Forensic Computer Science}}  \bibinfo{volume}{1} (\bibinfo{year}{2009}), \bibinfo{pages}{22--28}.
\newblock


\bibitem[Papernot et~al\mbox{.}(2017)]%
        {papernot2017practical}
\bibfield{author}{\bibinfo{person}{Nicolas Papernot}, \bibinfo{person}{Patrick McDaniel}, \bibinfo{person}{Ian Goodfellow}, \bibinfo{person}{Somesh Jha}, \bibinfo{person}{Z~Berkay Celik}, {and} \bibinfo{person}{Ananthram Swami}.} \bibinfo{year}{2017}\natexlab{}.
\newblock \showarticletitle{Practical black-box attacks against machine learning}. In \bibinfo{booktitle}{\emph{Proceedings of the 2017 ACM on Asia conference on computer and communications security}}. \bibinfo{pages}{506--519}.
\newblock


\bibitem[Park and Lee(2020)]%
        {park2020}
\bibfield{author}{\bibinfo{person}{Eunji Park} {and} \bibinfo{person}{Byungjoo Lee}.} \bibinfo{year}{2020}\natexlab{}.
\newblock \showarticletitle{An intermittent click planning model}. In \bibinfo{booktitle}{\emph{Proceedings of the 2020 CHI Conference on Human Factors in Computing Systems}}. \bibinfo{pages}{1--13}.
\newblock


\bibitem[Pei et~al\mbox{.}(2004)]%
        {pei2004mining}
\bibfield{author}{\bibinfo{person}{Jian Pei}, \bibinfo{person}{Jiawei Han}, \bibinfo{person}{Behzad Mortazavi-Asl}, \bibinfo{person}{Jianyong Wang}, \bibinfo{person}{Helen Pinto}, \bibinfo{person}{Qiming Chen}, \bibinfo{person}{Umeshwar Dayal}, {and} \bibinfo{person}{Mei-Chun Hsu}.} \bibinfo{year}{2004}\natexlab{}.
\newblock \showarticletitle{Mining sequential patterns by pattern-growth: The prefixspan approach}.
\newblock \bibinfo{journal}{\emph{IEEE Transactions on knowledge and data engineering}} \bibinfo{volume}{16}, \bibinfo{number}{11} (\bibinfo{year}{2004}), \bibinfo{pages}{1424--1440}.
\newblock


\bibitem[Phillips et~al\mbox{.}(2000)]%
        {phillips2000introduction}
\bibfield{author}{\bibinfo{person}{P~Jonathon Phillips}, \bibinfo{person}{Alvin Martin}, \bibinfo{person}{Charles~L Wilson}, {and} \bibinfo{person}{Mark Przybocki}.} \bibinfo{year}{2000}\natexlab{}.
\newblock \showarticletitle{An introduction evaluating biometric systems}.
\newblock \bibinfo{journal}{\emph{Computer}} \bibinfo{volume}{33}, \bibinfo{number}{2} (\bibinfo{year}{2000}), \bibinfo{pages}{56--63}.
\newblock


\bibitem[Pimenta et~al\mbox{.}(2015)]%
        {pimenta2015detection}
\bibfield{author}{\bibinfo{person}{Andr{\'e} Pimenta}, \bibinfo{person}{Davide Carneiro}, \bibinfo{person}{Paulo Novais}, {and} \bibinfo{person}{Jos{\'e} Neves}.} \bibinfo{year}{2015}\natexlab{}.
\newblock \showarticletitle{Detection of distraction and fatigue in groups through the analysis of interaction patterns with computers}.
\newblock In \bibinfo{booktitle}{\emph{Intelligent Distributed Computing VIII}}. \bibinfo{publisher}{Springer}, \bibinfo{pages}{29--39}.
\newblock


\bibitem[Pusara(2007)]%
        {pusara2007examination}
\bibfield{author}{\bibinfo{person}{Maja Pusara}.} \bibinfo{year}{2007}\natexlab{}.
\newblock \emph{\bibinfo{title}{An examination of user behavior for user re-authentication}}.
\newblock \bibinfo{thesistype}{Ph.\,D. Dissertation}. \bibinfo{school}{Purdue University}.
\newblock


\bibitem[Pusara and Brodley(2004)]%
        {pusara2004user}
\bibfield{author}{\bibinfo{person}{Maja Pusara} {and} \bibinfo{person}{Carla~E Brodley}.} \bibinfo{year}{2004}\natexlab{}.
\newblock \showarticletitle{User re-authentication via mouse movements}. In \bibinfo{booktitle}{\emph{Proceedings of the 2004 ACM workshop on Visualization and data mining for computer security}}. \bibinfo{pages}{1--8}.
\newblock


\bibitem[Ratha et~al\mbox{.}(2001)]%
        {ratha2001analysis}
\bibfield{author}{\bibinfo{person}{Nalini~K Ratha}, \bibinfo{person}{Jonathan~H Connell}, {and} \bibinfo{person}{Ruud~M Bolle}.} \bibinfo{year}{2001}\natexlab{}.
\newblock \showarticletitle{An analysis of minutiae matching strength}. In \bibinfo{booktitle}{\emph{International Conference on Audio-and Video-Based Biometric Person Authentication}}. Springer, \bibinfo{pages}{223--228}.
\newblock


\bibitem[Revett et~al\mbox{.}(2008)]%
        {revett2008survey}
\bibfield{author}{\bibinfo{person}{Kenneth Revett}, \bibinfo{person}{Hamid Jahankhani}, \bibinfo{person}{Sergio~Tenreiro De~Magalhaes}, {and} \bibinfo{person}{Henrique~MD Santos}.} \bibinfo{year}{2008}\natexlab{}.
\newblock \showarticletitle{A survey of user authentication based on mouse dynamics}. In \bibinfo{booktitle}{\emph{International Conference on Global e-Security}}. Springer, \bibinfo{pages}{210--219}.
\newblock


\bibitem[Ross and Jain(2003)]%
        {ross2003information}
\bibfield{author}{\bibinfo{person}{Arun Ross} {and} \bibinfo{person}{Anil Jain}.} \bibinfo{year}{2003}\natexlab{}.
\newblock \showarticletitle{Information fusion in biometrics}.
\newblock \bibinfo{journal}{\emph{Pattern recognition letters}} \bibinfo{volume}{24}, \bibinfo{number}{13} (\bibinfo{year}{2003}), \bibinfo{pages}{2115--2125}.
\newblock


\bibitem[Ross et~al\mbox{.}(2006)]%
        {ross2006handbook}
\bibfield{author}{\bibinfo{person}{Arun~A Ross}, \bibinfo{person}{Karthik Nandakumar}, {and} \bibinfo{person}{Anil~K Jain}.} \bibinfo{year}{2006}\natexlab{}.
\newblock \bibinfo{booktitle}{\emph{Handbook of multibiometrics}}. Vol.~\bibinfo{volume}{6}.
\newblock \bibinfo{publisher}{Springer Science \& Business Media}.
\newblock


\bibitem[Salman and Hameed(2018)]%
        {salman2018using}
\bibfield{author}{\bibinfo{person}{Osama~A Salman} {and} \bibinfo{person}{Sarab~M Hameed}.} \bibinfo{year}{2018}\natexlab{}.
\newblock \showarticletitle{Using mouse dynamics for continuous user authentication}. In \bibinfo{booktitle}{\emph{Proceedings of the Future Technologies Conference}}. Springer, \bibinfo{pages}{776--787}.
\newblock


\bibitem[Salthouse(1986)]%
        {salthouse1986perceptual}
\bibfield{author}{\bibinfo{person}{Timothy~A Salthouse}.} \bibinfo{year}{1986}\natexlab{}.
\newblock \showarticletitle{Perceptual, cognitive, and motoric aspects of transcription typing.}
\newblock \bibinfo{journal}{\emph{Psychological bulletin}} \bibinfo{volume}{99}, \bibinfo{number}{3} (\bibinfo{year}{1986}), \bibinfo{pages}{303}.
\newblock


\bibitem[Sargent and GreenLeaf({[n.\,d.]})]%
        {safe_lock_description}
\bibfield{author}{\bibinfo{person}{Sargent} {and} \bibinfo{person}{GreenLeaf}.} \bibinfo{year}{[n.\,d.]}\natexlab{}.
\newblock \bibinfo{title}{Four Wheel Safe Locks}.
\newblock
\newblock
\urldef\tempurl%
\url{https://classlocks.com.au/downloads/Safe Instructions/S&G 4 Wheel.pdf}
\showURL{%
\tempurl}


\bibitem[Sayed(2009)]%
        {sayed2009static}
\bibfield{author}{\bibinfo{person}{Bassam Sayed}.} \bibinfo{year}{2009}\natexlab{}.
\newblock \emph{\bibinfo{title}{A static authentication framework based on mouse gesture dynamics}}.
\newblock \bibinfo{thesistype}{Ph.\,D. Dissertation}.
\newblock


\bibitem[Sayed et~al\mbox{.}(2013)]%
        {sayed2013biometric}
\bibfield{author}{\bibinfo{person}{Bassam Sayed}, \bibinfo{person}{Issa Traor{\'e}}, \bibinfo{person}{Isaac Woungang}, {and} \bibinfo{person}{Mohammad~S Obaidat}.} \bibinfo{year}{2013}\natexlab{}.
\newblock \showarticletitle{Biometric authentication using mouse gesture dynamics}.
\newblock \bibinfo{journal}{\emph{IEEE systems journal}} \bibinfo{volume}{7}, \bibinfo{number}{2} (\bibinfo{year}{2013}), \bibinfo{pages}{262--274}.
\newblock


\bibitem[Sch{\"o}lkopf et~al\mbox{.}(2001)]%
        {scholkopf2001estimating}
\bibfield{author}{\bibinfo{person}{Bernhard Sch{\"o}lkopf}, \bibinfo{person}{John~C Platt}, \bibinfo{person}{John Shawe-Taylor}, \bibinfo{person}{Alex~J Smola}, {and} \bibinfo{person}{Robert~C Williamson}.} \bibinfo{year}{2001}\natexlab{}.
\newblock \showarticletitle{Estimating the support of a high-dimensional distribution}.
\newblock \bibinfo{journal}{\emph{Neural computation}} \bibinfo{volume}{13}, \bibinfo{number}{7} (\bibinfo{year}{2001}), \bibinfo{pages}{1443--1471}.
\newblock


\bibitem[Schonlau et~al\mbox{.}(2001)]%
        {schonlau2001computer}
\bibfield{author}{\bibinfo{person}{Matthias Schonlau}, \bibinfo{person}{William DuMouchel}, \bibinfo{person}{Wen-Hua Ju}, \bibinfo{person}{Alan~F Karr}, \bibinfo{person}{Martin Theus}, {and} \bibinfo{person}{Yehuda Vardi}.} \bibinfo{year}{2001}\natexlab{}.
\newblock \showarticletitle{Computer intrusion: Detecting masquerades}.
\newblock \bibinfo{journal}{\emph{Statistical science}} (\bibinfo{year}{2001}), \bibinfo{pages}{58--74}.
\newblock


\bibitem[Schulz(2006)]%
        {schulz2006mouse}
\bibfield{author}{\bibinfo{person}{Douglas~A Schulz}.} \bibinfo{year}{2006}\natexlab{}.
\newblock \showarticletitle{Mouse curve biometrics}. In \bibinfo{booktitle}{\emph{2006 Biometrics Symposium: Special Session on Research at the Biometric Consortium Conference}}. IEEE, \bibinfo{pages}{1--6}.
\newblock


\bibitem[Shannon and Weaver(1949)]%
        {shannon1949mathematical}
\bibfield{author}{\bibinfo{person}{Claude~E Shannon} {and} \bibinfo{person}{Warren Weaver}.} \bibinfo{year}{1949}\natexlab{}.
\newblock \showarticletitle{The mathematical theory of information}.
\newblock \bibinfo{journal}{\emph{Urbana: University of Illinois Press}}  \bibinfo{volume}{97} (\bibinfo{year}{1949}).
\newblock


\bibitem[Shen et~al\mbox{.}(2012a)]%
        {shen2012continuous}
\bibfield{author}{\bibinfo{person}{Chao Shen}, \bibinfo{person}{Zhongmin Cai}, {and} \bibinfo{person}{Xiaohong Guan}.} \bibinfo{year}{2012}\natexlab{a}.
\newblock \showarticletitle{Continuous authentication for mouse dynamics: A pattern-growth approach}. In \bibinfo{booktitle}{\emph{IEEE/IFIP International Conference on Dependable Systems and Networks (DSN 2012)}}. IEEE, \bibinfo{pages}{1--12}.
\newblock


\bibitem[Shen et~al\mbox{.}(2010)]%
        {shen2010hypo}
\bibfield{author}{\bibinfo{person}{Chao Shen}, \bibinfo{person}{Zhongmin Cai}, \bibinfo{person}{Xiaohong Guan}, {and} \bibinfo{person}{Jinpei Cai}.} \bibinfo{year}{2010}\natexlab{}.
\newblock \showarticletitle{A hypo-optimum feature selection strategy for mouse dynamics in continuous identity authentication and monitoring}. In \bibinfo{booktitle}{\emph{2010 IEEE International Conference on Information Theory and Information Security}}. IEEE, \bibinfo{pages}{349--353}.
\newblock


\bibitem[Shen et~al\mbox{.}(2012c)]%
        {shen2012user}
\bibfield{author}{\bibinfo{person}{Chao Shen}, \bibinfo{person}{Zhongmin Cai}, \bibinfo{person}{Xiaohong Guan}, \bibinfo{person}{Youtian Du}, {and} \bibinfo{person}{Roy~A Maxion}.} \bibinfo{year}{2012}\natexlab{c}.
\newblock \showarticletitle{User authentication through mouse dynamics}.
\newblock \bibinfo{journal}{\emph{IEEE Transactions on Information Forensics and Security}} \bibinfo{volume}{8}, \bibinfo{number}{1} (\bibinfo{year}{2012}), \bibinfo{pages}{16--30}.
\newblock


\bibitem[Shen et~al\mbox{.}(2014)]%
        {shen2014performance}
\bibfield{author}{\bibinfo{person}{Chao Shen}, \bibinfo{person}{Zhongmin Cai}, \bibinfo{person}{Xiaohong Guan}, {and} \bibinfo{person}{Roy Maxion}.} \bibinfo{year}{2014}\natexlab{}.
\newblock \showarticletitle{Performance evaluation of anomaly-detection algorithms for mouse dynamics}.
\newblock \bibinfo{journal}{\emph{computers \& security}}  \bibinfo{volume}{45} (\bibinfo{year}{2014}), \bibinfo{pages}{156--171}.
\newblock


\bibitem[Shen et~al\mbox{.}(2009)]%
        {shen2009feature}
\bibfield{author}{\bibinfo{person}{Chao Shen}, \bibinfo{person}{Zhongmin Cai}, \bibinfo{person}{Xiaohong Guan}, \bibinfo{person}{Huilan Sha}, {and} \bibinfo{person}{Jingzi Du}.} \bibinfo{year}{2009}\natexlab{}.
\newblock \showarticletitle{Feature analysis of mouse dynamics in identity authentication and monitoring}. In \bibinfo{booktitle}{\emph{2009 IEEE International Conference on Communications}}. IEEE, \bibinfo{pages}{1--5}.
\newblock


\bibitem[Shen et~al\mbox{.}(2012b)]%
        {shen2012effectiveness}
\bibfield{author}{\bibinfo{person}{Chao Shen}, \bibinfo{person}{Zhongmin Cai}, \bibinfo{person}{Xiaohong Guan}, {and} \bibinfo{person}{Jialin Wang}.} \bibinfo{year}{2012}\natexlab{b}.
\newblock \showarticletitle{On the effectiveness and applicability of mouse dynamics biometric for static authentication: A benchmark study}. In \bibinfo{booktitle}{\emph{2012 5th IAPR International Conference on Biometrics (ICB)}}. IEEE, \bibinfo{pages}{378--383}.
\newblock


\bibitem[Shen et~al\mbox{.}(2017)]%
        {shen2017pattern}
\bibfield{author}{\bibinfo{person}{Chao Shen}, \bibinfo{person}{Yufei Chen}, \bibinfo{person}{Xiaohong Guan}, {and} \bibinfo{person}{Roy~A Maxion}.} \bibinfo{year}{2017}\natexlab{}.
\newblock \showarticletitle{Pattern-growth based mining mouse-interaction behavior for an active user authentication system}.
\newblock \bibinfo{journal}{\emph{IEEE transactions on dependable and secure computing}} \bibinfo{volume}{17}, \bibinfo{number}{2} (\bibinfo{year}{2017}), \bibinfo{pages}{335--349}.
\newblock


\bibitem[Shi et~al\mbox{.}(2023)]%
        {shi2023user}
\bibfield{author}{\bibinfo{person}{Yutong Shi}, \bibinfo{person}{Xiujuan Wang}, \bibinfo{person}{Kangfeng Zheng}, {and} \bibinfo{person}{Siwei Cao}.} \bibinfo{year}{2023}\natexlab{}.
\newblock \showarticletitle{User authentication method based on keystroke dynamics and mouse dynamics using HDA}.
\newblock \bibinfo{journal}{\emph{Multimedia Systems}} \bibinfo{volume}{29}, \bibinfo{number}{2} (\bibinfo{year}{2023}), \bibinfo{pages}{653--668}.
\newblock


\bibitem[Shneiderman(1980)]%
        {shneiderman1980software}
\bibfield{author}{\bibinfo{person}{Ben Shneiderman}.} \bibinfo{year}{1980}\natexlab{}.
\newblock \bibinfo{booktitle}{\emph{Software psychology: Human factors in computer and information systems (Winthrop computer systems series)}}.
\newblock \bibinfo{publisher}{Winthrop Publishers}.
\newblock


\bibitem[Siddiqui et~al\mbox{.}(2021)]%
        {siddiqui2021continuous}
\bibfield{author}{\bibinfo{person}{Nyle Siddiqui}, \bibinfo{person}{Rushit Dave}, {and} \bibinfo{person}{Naeem Seliya}.} \bibinfo{year}{2021}\natexlab{}.
\newblock \showarticletitle{Continuous Authentication Using Mouse Movements, Machine Learning, and Minecraft}.
\newblock \bibinfo{journal}{\emph{arXiv preprint arXiv:2110.11080}} (\bibinfo{year}{2021}).
\newblock


\bibitem[Siddiqui et~al\mbox{.}(2022)]%
        {siddiqui2022}
\bibfield{author}{\bibinfo{person}{Nyle Siddiqui}, \bibinfo{person}{Rushit Dave}, \bibinfo{person}{Mounika Vanamala}, {and} \bibinfo{person}{Naeem Seliya}.} \bibinfo{year}{2022}\natexlab{}.
\newblock \showarticletitle{Machine and deep learning applications to mouse dynamics for continuous user authentication}.
\newblock \bibinfo{journal}{\emph{Machine Learning and Knowledge Extraction}} \bibinfo{volume}{4}, \bibinfo{number}{2} (\bibinfo{year}{2022}), \bibinfo{pages}{502--518}.
\newblock


\bibitem[Spillane(1975)]%
        {spillane1975keyboard}
\bibfield{author}{\bibinfo{person}{R Spillane}.} \bibinfo{year}{1975}\natexlab{}.
\newblock \showarticletitle{Keyboard apparatus for personal identification}.
\newblock \bibinfo{journal}{\emph{IBM Technical Disclosure Bulletin}}  \bibinfo{volume}{17} (\bibinfo{year}{1975}), \bibinfo{pages}{3346}.
\newblock


\bibitem[Syukri et~al\mbox{.}(1998)]%
        {syukri1998user}
\bibfield{author}{\bibinfo{person}{Agus~Fanar Syukri}, \bibinfo{person}{Eiji Okamoto}, {and} \bibinfo{person}{Masahiro Mambo}.} \bibinfo{year}{1998}\natexlab{}.
\newblock \showarticletitle{A user identification system using signature written with mouse}. In \bibinfo{booktitle}{\emph{Australasian conference on information security and privacy}}. Springer, \bibinfo{pages}{403--414}.
\newblock


\bibitem[Tan et~al\mbox{.}(2017)]%
        {tan2017insights}
\bibfield{author}{\bibinfo{person}{Yi~Xiang~Marcus Tan}, \bibinfo{person}{Alexander Binder}, {and} \bibinfo{person}{Arunava Roy}.} \bibinfo{year}{2017}\natexlab{}.
\newblock \showarticletitle{Insights from curve fitting models in mouse dynamics authentication systems}. In \bibinfo{booktitle}{\emph{2017 IEEE Conference on Application, Information and Network Security (AINS)}}. IEEE, \bibinfo{pages}{42--47}.
\newblock


\bibitem[Tan et~al\mbox{.}(2019)]%
        {tan2019adversarial}
\bibfield{author}{\bibinfo{person}{Yi~Xiang~Marcus Tan}, \bibinfo{person}{Alfonso Iacovazzi}, \bibinfo{person}{Ivan Homoliak}, \bibinfo{person}{Yuval Elovici}, {and} \bibinfo{person}{Alexander Binder}.} \bibinfo{year}{2019}\natexlab{}.
\newblock \showarticletitle{Adversarial attacks on remote user authentication using behavioural mouse dynamics}. In \bibinfo{booktitle}{\emph{2019 International Joint Conference on Neural Networks (IJCNN)}}. IEEE, \bibinfo{pages}{1--10}.
\newblock


\bibitem[Teh et~al\mbox{.}(2013)]%
        {teh2013survey}
\bibfield{author}{\bibinfo{person}{Pin~Shen Teh}, \bibinfo{person}{Andrew Beng~Jin Teoh}, {and} \bibinfo{person}{Shigang Yue}.} \bibinfo{year}{2013}\natexlab{}.
\newblock \showarticletitle{A survey of keystroke dynamics biometrics}.
\newblock \bibinfo{journal}{\emph{The Scientific World Journal}}  \bibinfo{volume}{2013} (\bibinfo{year}{2013}).
\newblock


\bibitem[Theodoridis et~al\mbox{.}(2010)]%
        {theodoridis2010introduction}
\bibfield{author}{\bibinfo{person}{Sergios Theodoridis}, \bibinfo{person}{Aggelos Pikrakis}, \bibinfo{person}{Konstantinos Koutroumbas}, {and} \bibinfo{person}{Dionisis Cavouras}.} \bibinfo{year}{2010}\natexlab{}.
\newblock \bibinfo{booktitle}{\emph{Introduction to pattern recognition: a matlab approach}}.
\newblock \bibinfo{publisher}{Academic Press}.
\newblock


\bibitem[TN and Pramod(2023)]%
        {tn2023insider}
\bibfield{author}{\bibinfo{person}{Nisha TN} {and} \bibinfo{person}{Dhanya Pramod}.} \bibinfo{year}{2023}\natexlab{}.
\newblock \showarticletitle{Insider Intrusion Detection Techniques: A State-of-the-Art Review}.
\newblock \bibinfo{journal}{\emph{Journal of Computer Information Systems}} (\bibinfo{year}{2023}), \bibinfo{pages}{1--18}.
\newblock


\bibitem[Uludag and Jain(2004)]%
        {uludag2004attacks}
\bibfield{author}{\bibinfo{person}{Umut Uludag} {and} \bibinfo{person}{Anil~K Jain}.} \bibinfo{year}{2004}\natexlab{}.
\newblock \showarticletitle{Attacks on biometric systems: a case study in fingerprints}. In \bibinfo{booktitle}{\emph{Security, steganography, and watermarking of multimedia contents VI}}, Vol.~\bibinfo{volume}{5306}. International Society for Optics and Photonics, \bibinfo{pages}{622--633}.
\newblock


\bibitem[Unknown(2021)]%
        {calculus_book}
\bibfield{author}{\bibinfo{person}{Unknown}.} \bibinfo{year}{2021}\natexlab{}.
\newblock \bibinfo{title}{Applications of the Derivative}.
\newblock
\newblock
\urldef\tempurl%
\url{https://understandingcalculus.com/chapters/06/6-2.php}
\showURL{%
\tempurl}


\bibitem[Vacca(2007)]%
        {vacca2007biometric}
\bibfield{author}{\bibinfo{person}{John~R Vacca}.} \bibinfo{year}{2007}\natexlab{}.
\newblock \bibinfo{booktitle}{\emph{Biometric technologies and verification systems}}.
\newblock \bibinfo{publisher}{Elsevier}.
\newblock


\bibitem[W3Schools({[n.\,d.]})]%
        {detla}
\bibfield{author}{\bibinfo{person}{W3Schools}.} \bibinfo{year}{[n.\,d.]}\natexlab{}.
\newblock \bibinfo{title}{MouseWheel-Deltay}.
\newblock
\newblock
\urldef\tempurl%
\url{https://www.w3schools.com/jsref/event_wheel_deltay.asp}
\showURL{%
\tempurl}


\bibitem[Wahab et~al\mbox{.}(2023)]%
        {codaspy-usability}
\bibfield{author}{\bibinfo{person}{Ahmed~Anu Wahab}, \bibinfo{person}{Daqing Hou}, {and} \bibinfo{person}{Stephanie Schuckers}.} \bibinfo{year}{2023}\natexlab{}.
\newblock \showarticletitle{A User Study of Keystroke Dynamics as Second Factor in Web MFA}. In \bibinfo{booktitle}{\emph{Proceedings of the Thirteenth ACM Conference on Data and Application Security and Privacy}} (Charlotte, NC, USA) \emph{(\bibinfo{series}{CODASPY '23})}. \bibinfo{publisher}{Association for Computing Machinery}, \bibinfo{address}{New York, NY, USA}, \bibinfo{pages}{61–72}.
\newblock
\showISBNx{9798400700675}
\urldef\tempurl%
\url{https://doi.org/10.1145/3577923.3583642}
\showDOI{\tempurl}


\bibitem[William and Harter(1899)]%
        {william1899studies}
\bibfield{author}{\bibinfo{person}{Lowe~Bryan William} {and} \bibinfo{person}{Noble Harter}.} \bibinfo{year}{1899}\natexlab{}.
\newblock \showarticletitle{Studies on the telegraphic language: The acquisition of a hierarchy of habits.}
\newblock \bibinfo{journal}{\emph{Psychological review}} \bibinfo{volume}{6}, \bibinfo{number}{4} (\bibinfo{year}{1899}), \bibinfo{pages}{345}.
\newblock


\bibitem[Williams and Zipser(1989)]%
        {williams1989learning}
\bibfield{author}{\bibinfo{person}{Ronald~J Williams} {and} \bibinfo{person}{David Zipser}.} \bibinfo{year}{1989}\natexlab{}.
\newblock \showarticletitle{A learning algorithm for continually running fully recurrent neural networks}.
\newblock \bibinfo{journal}{\emph{Neural computation}} \bibinfo{volume}{1}, \bibinfo{number}{2} (\bibinfo{year}{1989}), \bibinfo{pages}{270--280}.
\newblock


\bibitem[Wobbrock et~al\mbox{.}(2008)]%
        {wobbrock2008}
\bibfield{author}{\bibinfo{person}{Jacob~O Wobbrock}, \bibinfo{person}{Edward Cutrell}, \bibinfo{person}{Susumu Harada}, {and} \bibinfo{person}{I~Scott MacKenzie}.} \bibinfo{year}{2008}\natexlab{}.
\newblock \showarticletitle{An error model for pointing based on Fitts' law}. In \bibinfo{booktitle}{\emph{Proceedings of the SIGCHI conference on human factors in computing systems}}. \bibinfo{pages}{1613--1622}.
\newblock


\bibitem[Wondracek et~al\mbox{.}(2010)]%
        {wondracek2010practical}
\bibfield{author}{\bibinfo{person}{Gilbert Wondracek}, \bibinfo{person}{Thorsten Holz}, \bibinfo{person}{Engin Kirda}, {and} \bibinfo{person}{Christopher Kruegel}.} \bibinfo{year}{2010}\natexlab{}.
\newblock \showarticletitle{A practical attack to de-anonymize social network users}. In \bibinfo{booktitle}{\emph{2010 ieee symposium on security and privacy}}. IEEE, \bibinfo{pages}{223--238}.
\newblock


\bibitem[Wu and Liu(2008)]%
        {wu2008queuing}
\bibfield{author}{\bibinfo{person}{Changxu Wu} {and} \bibinfo{person}{Yili Liu}.} \bibinfo{year}{2008}\natexlab{}.
\newblock \showarticletitle{Queuing network modeling of transcription typing}.
\newblock \bibinfo{journal}{\emph{ACM Transactions on Computer-Human Interaction (TOCHI)}} \bibinfo{volume}{15}, \bibinfo{number}{1} (\bibinfo{year}{2008}), \bibinfo{pages}{1--45}.
\newblock


\bibitem[Yampolskiy and Govindaraju(2008)]%
        {yampolskiy2008behavioural}
\bibfield{author}{\bibinfo{person}{Roman~V Yampolskiy} {and} \bibinfo{person}{Venu Govindaraju}.} \bibinfo{year}{2008}\natexlab{}.
\newblock \showarticletitle{Behavioural biometrics: a survey and classification}.
\newblock \bibinfo{journal}{\emph{International Journal of Biometrics}} \bibinfo{volume}{1}, \bibinfo{number}{1} (\bibinfo{year}{2008}), \bibinfo{pages}{81--113}.
\newblock


\bibitem[Zhai et~al\mbox{.}(2004)]%
        {zhai2004}
\bibfield{author}{\bibinfo{person}{Shumin Zhai}, \bibinfo{person}{Jing Kong}, {and} \bibinfo{person}{Xiangshi Ren}.} \bibinfo{year}{2004}\natexlab{}.
\newblock \showarticletitle{Speed--accuracy tradeoff in Fitts’ law tasks—on the equivalency of actual and nominal pointing precision}.
\newblock \bibinfo{journal}{\emph{International journal of human-computer studies}} \bibinfo{volume}{61}, \bibinfo{number}{6} (\bibinfo{year}{2004}), \bibinfo{pages}{823--856}.
\newblock


\bibitem[Zheng(2014)]%
        {zheng2014exploiting}
\bibfield{author}{\bibinfo{person}{Nan Zheng}.} \bibinfo{year}{2014}\natexlab{}.
\newblock \showarticletitle{Exploiting behavioral biometrics for user security enhancements}.
\newblock  (\bibinfo{year}{2014}).
\newblock


\bibitem[Zheng et~al\mbox{.}(2011)]%
        {zheng2011efficient}
\bibfield{author}{\bibinfo{person}{Nan Zheng}, \bibinfo{person}{Aaron Paloski}, {and} \bibinfo{person}{Haining Wang}.} \bibinfo{year}{2011}\natexlab{}.
\newblock \showarticletitle{An efficient user verification system via mouse movements}. In \bibinfo{booktitle}{\emph{Proceedings of the 18th ACM conference on Computer and communications security}}. \bibinfo{pages}{139--150}.
\newblock


\bibitem[Zheng et~al\mbox{.}(2016)]%
        {zheng2016efficient}
\bibfield{author}{\bibinfo{person}{Nan Zheng}, \bibinfo{person}{Aaron Paloski}, {and} \bibinfo{person}{Haining Wang}.} \bibinfo{year}{2016}\natexlab{}.
\newblock \showarticletitle{An efficient user verification system using angle-based mouse movement biometrics}.
\newblock \bibinfo{journal}{\emph{ACM Transactions on Information and System Security (TISSEC)}} \bibinfo{volume}{18}, \bibinfo{number}{3} (\bibinfo{year}{2016}), \bibinfo{pages}{1--27}.
\newblock


\bibitem[Zhou et~al\mbox{.}(2009)]%
        {zhou2009}
\bibfield{author}{\bibinfo{person}{Xiaolei Zhou}, \bibinfo{person}{Xiang Cao}, {and} \bibinfo{person}{Xiangshi Ren}.} \bibinfo{year}{2009}\natexlab{}.
\newblock \showarticletitle{Speed-accuracy tradeoff in trajectory-based tasks with temporal constraint}. In \bibinfo{booktitle}{\emph{Human-Computer Interaction--INTERACT 2009: 12th IFIP TC 13 International Conference, Uppsala, Sweden, August 24-28, 2009, Proceedings, Part I 12}}. Springer, \bibinfo{pages}{906--919}.
\newblock


\end{thebibliography}
\end{document}